\documentclass{aa}

\usepackage{graphicx}
\usepackage{txfonts}
\usepackage{placeins}
\usepackage{natbib}    
\bibpunct{(}{)}{;}{a}{}{,}
\usepackage{upgreek}
\usepackage{multirow}

\usepackage{hyperref}
\usepackage{cleveref}
\crefname{section}{Sect.}{Sects.}
\Crefname{section}{Section}{Sections}
\crefname{figure}{Fig.}{Figs.}
\Crefname{figure}{Figure}{Figures}
\crefname{table}{Table}{Tables}
\crefname{equation}{Eq.}{Eqs.}
\Crefname{equation}{Equation}{Equations}

\newcommand{\dd}{\mathrm{d}}
\newcommand{\dv}[2]{\frac{\dd #1}{\dd #2}}
\newcommand{\pdv}[2]{\frac{\partial #1}{\partial #2}}
\newcommand{\order}[1]{\mathcal{O}\left(#1\right)}

\newcommand{\noteadded}[1]{%
  \par\addvspace{6pt}%
  \begingroup
  \tiny
  \noindent\textit{Note added in proof.}\hskip.5em #1\par
  \endgroup
  \addvspace{6pt}%
}

\begin{document}

\title{Cooling, conduction, and compact objects: Gravothermal evolution of dissipative self-interacting dark matter halos}

\titlerunning{Gravothermal evolution of dissipative SIDM halos}

\subtitle{}

   \author{Ludwig D.~Schmidt\inst{\ref{inst:tum}}\fnmsep\thanks{E-mail: ludwig.schmidt@tum.de}
        \and Moritz S.~Fischer\inst{\ref{inst:dipc},\ref{inst:usm}}
        \and Mathias Garny\inst{\ref{inst:tum}}
        }

   \institute{Technische Universität München, TUM School of Natural Sciences, Physik Department T31, James-Franck-Straße 1, D-85748 Garching, Germany\label{inst:tum}
    \and
    Donostia International Physics Center (DIPC), Paseo Manuel de Lardizabal 4, 20018 Donostia-San Sebastian, Spain\label{inst:dipc}
    \and
    Universitäts-Sternwarte, Fakultät für Physik, Ludwig-Maximilians-Universität München, Scheinerstr.\ 1, D-81679 München, Germany\label{inst:usm}
   }

   \date{Received 9 July 2026 / Accepted 9 September 2026}
 
  \abstract
   {Many proposed self-interacting dark matter (SIDM) models yield radiative processes that can dissipate energy. Understanding their impact on astrophysical objects through simulations and comparing the results with observations may thus constrain SIDM models.}
   {We systematically investigated how dissipation alters the gravothermal evolution of isolated SIDM halos by independently varying dissipation and heat conduction and identified potential observational signatures.} 
   {We present the first extension of the \(N\)-body formalism for frequent small-angle self-interactions to include effective dissipation. We compared all results for isolated halos with a dissipative gravothermal fluid model to assess the validity and limitations of the latter.}
   {We find that dissipation qualitatively changes the gravothermal evolution of SIDM halos beyond simply accelerating collapse. Sufficiently strong central cooling can invert the usual role of heat conduction: the formation of an isothermal core is suppressed such that conduction remains directed inward throughout the evolution. Outer halo regions beyond the scale radius can cool efficiently rather than being heated by conduction, resulting in a larger region of mass infall and a less pronounced indentation between the core and the outer halo in the final density profile. These effects depend strongly on the cooling rate but are comparatively insensitive to the angular dependence of the self-interaction cross section. We further show that weakly dissipative self-interactions can explain the properties of the recently observed strong lens perturber in JVAS~B1938+666 with significantly shorter evolution times or, equivalently, smaller cross sections compared to the elastic case.}
   {Our results open a new route to connecting halo structure and recently reported compact objects to dark-sector microphysics.}


   \maketitle
   \nolinenumbers

\section{Introduction}
Self-interacting dark matter (SIDM) has emerged as a compelling extension of the cold dark matter (CDM) paradigm \citep{Spergel2000,Tulin2018,Adhikari2025}. It is well motivated from a particle-physics perspective and addresses several small-scale challenges of the \(\Lambda\)CDM model (see \citealt{Bullock2017} for a review). Self-interactions enable collisional heat conduction, which causes a gravothermal evolution of dark matter (DM) halos with initial core formation and subsequent collapse \citep{Balberg2002,Koda2011}. The emerging time dependence of the central density alleviates both the core-cusp problem and the too-big-to-fail problem \citep{Elbert2015}. 

Models of SIDM often invoke light mediators to achieve sizable cross sections at dwarf-galaxy velocities \citep[e.g.,][]{Tulin2013} while satisfying constraints at cluster scales \citep[e.g.,][]{Randall2008,Sagunski2021,Andrade2022}. Recently, \citet{Yu2026} collected different indications of compact objects that SIDM could explain. Most prominently, a dark million-solar-mass object in the JVAS~B1938+666 system at a redshift of \(z = 0.881\) has been identified via gravitational lensing \citep{Powell2025,Vegetti2026}. Its inferred mass distribution -- an unresolved central point mass embedded in a nearly constant surface density -- appears to be consistent with a collapsing SIDM halo. However, \citet{Vegetti2026} inferred that core collapse by this redshift requires a velocity-averaged elastic cross section as large as \(800\,\mathrm{cm}^2\,\mathrm{g}^{-1}\). This implies that the cross section would have to increase very strongly toward low velocities.

An alternative explanation is that dissipation accelerates gravothermal evolution \citep{Essig2019}, potentially allowing halos to enter the collapse phase before \(z = 0.881\), even for moderate self-interaction cross sections. This mechanism has also been proposed to explain the early formation of supermassive black holes \citep{Choquette2019,Xiao2021}.

Light mediators (e.g., dark photons) naturally permit dissipative processes via bremsstrahlung \citep{LankesterBroche2026} if allowed kinematically. Dissipation can become particularly efficient when DM bound states are present, which can be collisionally excited and subsequently decay via mediator emission -- closely analogous to radiative cooling in baryonic hydrogen during structure formation \citep{Spitzer1978}. Models featuring such bound states include mirror DM (\citealt{Foot2004}),  asymmetric DM \citep{Wise2014,Gresham2018}, dark quantum chromodynamics \citep{Kribs2016}, and atomic DM \citep{Kaplan2010}.

Using dissipative DM models, \citet{Foot2013} and \citet{Randall2015} attempted to explain the plane of satellite galaxies around the Andromeda Galaxy (M31). As first pointed out by \citet{Fan2013}, a dissipative subcomponent of DM can form a second disk (so-called double-disk DM). In its plasma phase, atomic DM may resolve some small-scale problems and lead to direct-detection signals \citep{Foot2015,Foot2015a,Foot2016}. Recent studies further suggest that atomic DM can create compact objects within DM halos \citep{Gurian2022}. Owing to its similarity to ordinary baryonic gas, the dissipative component is commonly simulated as a fluid \citep{Roy2023,Roy2025}.

However, this fluid approximation breaks down for less strongly self-interacting DM. Previous dissipative \(N\)-body simulations of DM typically assume an isotropic cross section and dissipate either a fixed amount of energy with a certain probability above a velocity threshold \citep{Huo2020} or a fixed fraction of the kinetic energy \citep{Shen2021}. However, realistic particle-physics cross sections, especially those mediated by light particles, are often strongly forward-peaked, rendering direct sampling computationally expensive \citep{Robertson2017}. To address this, \citet{Fischer2021} introduced an \(N\)-body method called fSIDM (frequently self-interacting dark matter), which treats frequent small-angle scatterings effectively as a drag force and diffusion acting on the simulation particles. We extended this framework to include dissipation and applied it to isolated DM halos.

Using a simple velocity-independent parametrization, we systematically investigated how dissipative self-interactions modify the gravothermal evolution of isolated DM halos by varying cooling and heat conduction independently. We compared the results with those of the gravothermal fluid model to test how well it reproduces the N-body evolution, and with an isotropic implementation analogous to \citet{Shen2021} to examine the influence of the cross section's angular dependence. Finally, we derived projected quantities accessible to gravitational lensing observations to identify potential observational signatures of dissipative SIDM. We further investigated whether dissipative SIDM can generate the inferred mass profile of the strong lens perturber in JVAS~B1938+666 \citep{Vegetti2026}.  

This paper is structured as follows. In \cref{sec:Method} we review the fSIDM method and extend it to include dissipation. We present our simulation results for isolated halos in \cref{sec:Results}, exploring variations in dissipation and heat conduction. In \cref{sec:Discussion} we interpret these results and compute projected quantities to compare them with lensing observations.

\section{Numerical method}\label{sec:Method}
In this section, we extend the effective fSIDM method introduced by \citet{Fischer2021} to incorporate dissipative interactions. The implementation is part of the public code \textsc{OpenGadget3}\footnote{\url{https://gitlab.lrz.de/AstroCodes/OpenGadget3}} \citep{Ragagnin2016,Fischer2024,Fischer2026,Dolag2026}, a descendant of \textsc{GADGET-2} \citep{Springel2005}. Throughout this work, bold symbols, \(\boldsymbol{v}\), denote 3-vectors, \(v\) the corresponding magnitude, and \(\hat{\boldsymbol{v}} = \boldsymbol{v}/v\) the corresponding unit vector. 

\subsection{Review of the elastic fSIDM method}\label{sec:fSIDMreview}
In the \(N\)-body framework, each simulation particle represents an ensemble of physical DM particles with a coarse-grained phase-space distribution. If two simulation particles, \(i\) and \(j\), of equal mass \(m\) interact at a relative velocity \(\Delta\boldsymbol{v}_{ij} = \boldsymbol{v}_i - \boldsymbol{v}_j\), frequent small-angle scatterings between the underlying physical particles give rise to an effective drag force, 
\begin{equation}\label{eq:dragForce}
    F_\mathrm{drag} = \frac{1}{2}\Delta v_{ij}^2\frac{\sigma_\mathrm{T}}{m_\chi}\int \mathrm{d}V \rho_i \rho_j,
\end{equation}
as derived by \citet{Fischer2021}, building on results from \citet{Kahlhoefer2014}.
This expression assumes that the numerical particles have spatially extended density distributions, \(\rho_i\) and \(\rho_j\), which are given by spline kernels of sizes \(h_i\) and \(h_j\) set by the number of neighboring particles. Since the DM particles are assumed to be indistinguishable, the modified momentum transfer cross section,
\begin{equation}\label{eq:TransferCrossSection}
    \sigma_\mathrm{T} = 2 \int\limits_0^1 \mathrm{d}\!\cos\theta\,(1-\cos\theta)\frac{\mathrm{d}\sigma}{\mathrm{d}\!\cos\theta}
\end{equation}
enters, with the scattering angle \(\theta\) measured in the center-of-mass system (CMS).

The algorithm proceeds in two steps, as illustrated by \cref{fig:Scattering}. First, a drag-induced velocity change,
\begin{equation}\label{eq:dragVelocity}
    \Delta \boldsymbol{v}_\mathrm{drag} = \frac{F_\mathrm{drag}}{m} \Delta t \Delta\hat{\boldsymbol{v}}_{ij}
,\end{equation}
is applied along the direction of the relative velocity. Second, a stochastic velocity kick,
\begin{equation}\label{eq:kickVelocity}
    \Delta v_\mathrm{rand}^2 = \Delta v_\mathrm{drag}(\Delta v_{ij} - \Delta v_\mathrm{drag})
,\end{equation} 
perpendicular to the relative motion restores the energy lost through drag and models diffusion. Overall, the velocities of all overlapping particles are updated according to
\begin{equation}\label{eq:velocityUpdate}
    \boldsymbol{v}'_i = \boldsymbol{v}_i - \Delta \boldsymbol{v}_\mathrm{drag} + \Delta \boldsymbol{v}_\mathrm{rand},\quad \boldsymbol{v}'_j = \boldsymbol{v}_j + \Delta \boldsymbol{v}_\mathrm{drag} - \Delta \boldsymbol{v}_\mathrm{rand},
\end{equation}
which ensures momentum conservation. The time step (\(\Delta t\)) was chosen such that the relative change in velocity due to the drag force remains smaller than a fixed constant \(\tau \ll 1\), i.e., \(\Delta v_\mathrm{drag}/\Delta v_{ij} < \tau\) \citep{Fischer2024}. For more details on the scheme for frequent self-interactions, we refer the reader to \citet{Fischer2021,Fischer2026}. 

\subsection{Dissipative extension}\label{sec:DissipativeExtension}
We next derive how \(\Delta \boldsymbol{v}_\mathrm{drag}\) and \(\Delta \boldsymbol{v}_\mathrm{rand}\) must change in a dissipative situation. As we will show, dissipation increases the drag. The corresponding energy loss is then not fully restored, such that the total energy of the \(N\)-body system decreases as desired. 

\begin{figure}[t]
    \centering
    \includegraphics[width=\hsize]{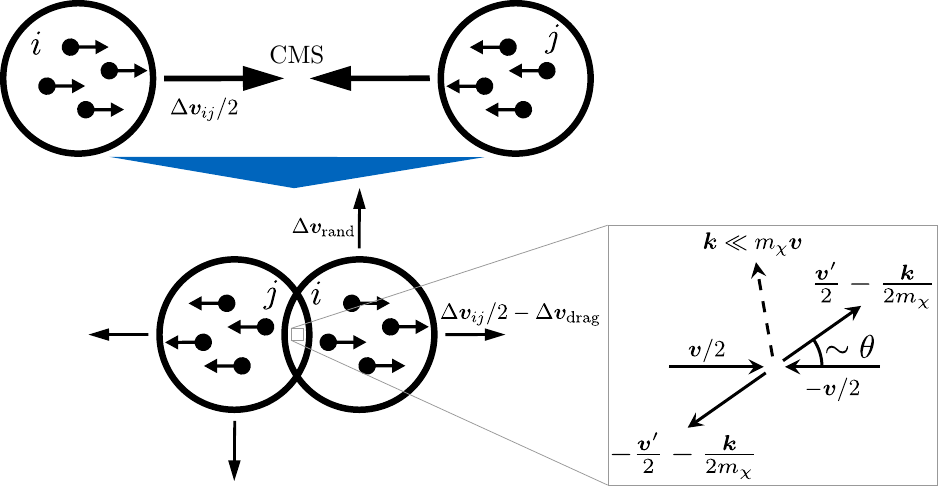}
    \caption{\textit{Left:} Schematic scattering of numerical particles. Both decelerate by \(\Delta \boldsymbol{v}_\mathrm{drag}\) and experience opposite \(\Delta \boldsymbol{v}_\mathrm{rand}\) in a random direction perpendicular to the relative velocity, \(\Delta \boldsymbol{v}_{ij}\). 
    \textit{Right:} Zoom-in to the level of physical particles for a dissipative process like that described by \cref{eq:dissipativeReation}. The physical particles undergo many such scatterings in a single numerical interaction. Therefore, the radiated momenta (\(\boldsymbol{k}\)) average out and the transverse momenta of the outgoing DM particles give rise to diffusion with vanishing drift in the CMS frame modeled by \(\Delta \boldsymbol{v}_\mathrm{rand}\). The radiated energy (\(k^0\)) accumulates and leads to an increased drag. For readability, the scattering angle and the magnitude of the radiated momentum (\(\boldsymbol{k}\)) are highly exaggerated.}
    \label{fig:Scattering}
\end{figure}

A very general radiative scattering process that can describe both bremsstrahlung and the collisional excitation of a short-lived state decaying under emission of radiation is given by
\begin{equation}\label{eq:dissipativeReation}
    \chi\left(\frac{m_\chi\boldsymbol{v}}{2}\right) + \chi\left(-\frac{m_\chi\boldsymbol{v}}{2}\right) \to \chi\left(\frac{m_\chi\boldsymbol{v}'}{2} - \frac{\boldsymbol{k}}{2}\right) + \chi\left(-\frac{m_\chi\boldsymbol{v}'}{2} - \frac{\boldsymbol{k}}{2}\right) + \phi(\boldsymbol{k})
\end{equation}
and illustrated in \cref{fig:Scattering} (right inset). The arguments denote the respective nonrelativistic three-momenta in the CMS frame.  One or more particles, collectively denoted by \(\phi\), carry away a total energy \(k^0\) and are not tracked in the simulation. We must therefore require that they are not significantly reabsorbed by DM particles. Furthermore, their mass must be sufficiently low to be produced on shell from the nonrelativistic kinetic energy. Intermediate excited states do not have to be included explicitly but enter via their impact on the differential cross section.

Using \(|\boldsymbol{k}| < k^0 \sim\order{v^2}\), we obtain
\begin{equation}
    k^0 = \frac{m_\chi}{4}(v^2 - v'^2) + \order{v^4},
\end{equation}
which implies \(k^0 \ll m_\chi v^2\) in the forward-scattering limit \(\boldsymbol{v}\sim\boldsymbol{v}'\) relevant for fSIDM. Using this to expand in \(k^0/(m_\chi v^2) \ll 1\) gives rise to
\begin{equation}
    v' = v\left[1 - \frac{2k^0}{m_\chi v^2} + \order{\left(\frac{k^0}{m_\chi v^2}\right)^2} \right].
\end{equation}
Following \citet{Kahlhoefer2014}, we computed the magnitude of the velocity change
 as\begin{equation}
    \delta v = \left|\frac{\boldsymbol{v}'}{2} - \frac{\boldsymbol{k}}{2m_\chi} - \frac{\boldsymbol{v}}{2}\right| \simeq \frac{|\boldsymbol{v}' - \boldsymbol{v}|}{2} \simeq v\left(1 - \frac{k^0}{m_\chi v^2}\right)\sin\frac{\theta}{2}
\end{equation}
and its component parallel to the initial relative velocity as
\begin{equation}
\begin{split}
    \delta v_\parallel &= \left(\frac{\boldsymbol{v}'}{2} - \frac{\boldsymbol{k}}{2m_\chi} - \frac{\boldsymbol{v}}{2}\right)\cdot \hat{\boldsymbol{v}} = \frac{v'}{2}\cos\theta - \frac{|\boldsymbol{k}|}{2m_\chi}\cos\theta_k - \frac{v}{2}\\
    &\simeq -v \left(\sin^2\frac{\theta}{2} + \frac{k^0}{m_\chi v^2}\cos\theta\right)
\end{split}
,\end{equation}
with \(\cos\theta \equiv \hat{\boldsymbol{v}}\cdot\hat{\boldsymbol{v}}'\) and \(\cos\theta_k \equiv \hat{\boldsymbol{v}}\cdot\hat{\boldsymbol{k}}\). Terms proportional to \(\boldsymbol{k}/m_\chi\) are subdominant compared to the velocities. They further involve \(\cos \theta_k\), which averages to zero over many scatterings if the emission of \(\phi\) is forward-backward symmetric. We therefore neglect these terms hereafter. 

For \(k^0 = 0\), the elastic expressions from \citet[Appendix A]{Kahlhoefer2014} are recovered. For small scattering angles, \(\theta\), and low levels of dissipation, \(k^0\), the hierarchy \(\delta v \gg \delta v_\parallel\) persists, implying that most momentum transfer occurs perpendicular to the initial direction. As the perpendicular components point in random directions within a plane, they average to zero over many scatterings. However, an average diffusion \(\langle\delta v_\perp^2\rangle\) in the perpendicular direction remains. At the same order in \(\theta\), the parallel component \(\langle\delta v_\parallel\rangle\) accumulates and gives rise to an effective drag force acting on the numerical particles. Assuming that all physical particles collide at the relative velocity, \(v = \Delta v_{ij}\), of the simulation particles, \(i\) and \(j\), the accumulated parallel momentum change is given by
\begin{equation}
    \dd p_\parallel = -F_\mathrm{drag}\Delta t = \int \dd{C} \dd{V} \rho_i \delta v_\parallel \equiv 
    \frac{1}{2}\frac{\sigma_\mathrm{T}}{m_\chi} r_\mathrm{diss} \Delta v_{ij}^2\Delta t\int \dd{V} \rho_i\rho_j.
\end{equation}
Here, we first integrated over the collisions, 
\begin{equation}
    \dd{C} = \frac{\rho_j}{m_\chi} \Delta v_{ij}\frac{\dd{\sigma}}{\dd{\Omega}\dd{k^0}}\dd{\Omega}\dd {k^0} \Delta t
,\end{equation}
of a single particle of \(i\) with the particles of \(j\), and then over all particles of \(i\). We defined the dissipation parameter, 
\begin{equation}\label{eq:rdiss}
    r_\mathrm{diss} \equiv 1 + \frac{8\uppi}{\sigma_\mathrm{T}}\int\limits_{m_\phi}^{m_\chi\Delta v_{ij}^2/4}\dd{k^0}\int\limits_0^1\dd{\cos\theta} \frac{\dd{\sigma}}{\dd{\Omega}\dd{k^0}} \frac{k^0}{m_\chi \Delta v_{ij}^2}\cos\theta
,\end{equation}
as the factor by which dissipation enhances the drag force relative to \cref{eq:dragForce}. The full expression for the drag velocity kick with dissipation thus reads
\begin{equation}\label{eq:dragVelDiss}
    \Delta \boldsymbol{v}_\mathrm{drag} = \Delta \hat{\boldsymbol{v}}_{ij} \frac{\Delta t}{2m} \Delta v_{ij}^2 \frac{\sigma_\mathrm{T}}{m_\chi} r_\mathrm{diss} \int \mathrm{d}V\,\rho_i\rho_j.
\end{equation}

In contrast to elastic scatterings, the kinetic energy of the numerical particles should not be conserved but decrease by 
\begin{equation}
    \Delta E = -\int \dd{C} \dd{V} \frac{\rho_i}{m_\chi} k^0 \simeq - \frac{1}{2} \Delta v_{ij}^3 \frac{\sigma_\mathrm{T}}{m_\chi}(r_\mathrm{diss} - 1) \Delta t \int \dd{V} \rho_i\rho_j
\end{equation}
in each time step, \(\Delta t\). Recovering \(r_\mathrm{diss}\) required introducing a factor \(\cos\theta\). This does not modify the angular integral if the cross section is dominated by small-angle scattering. If the recoil from the emitted radiation averages out over many interactions, i.e., if the cross section is symmetric under \(\boldsymbol{k}\to-\boldsymbol{k}\), the momentum-conserving velocity update given by \cref{eq:velocityUpdate} remains valid. Energy conservation 
\begin{equation}
    \Delta E = E'_\mathrm{kin} - E_\mathrm{kin} = \frac{m}{2}(\boldsymbol{v}_i'^2 + \boldsymbol{v}_j'^2) - \frac{m}{2}(\boldsymbol{v}_i^2 + \boldsymbol{v}_j^2)
\end{equation}
then implies a slightly modified velocity kick:
\begin{equation}\label{eq:RandKick}
    \Delta v_\mathrm{rand}^2 = \Delta v_\mathrm{drag}\left(\frac{\Delta v_{ij}}{r_\mathrm{diss}} - \Delta v_\mathrm{drag}\right),
\end{equation}
where \(\Delta v_\mathrm{drag}\) refers to \cref{eq:dragVelDiss}. 

To ensure \(\Delta v_\mathrm{drag}/\Delta v_{ij} < \tau \ll 1\) as before, the usual time step criterion of elastic fSIDM must be scaled down as \(\Delta t_\mathrm{diss} = \Delta t_\mathrm{el}/r_\mathrm{diss}\). For large \(r_\mathrm{diss}\), the requirement of positivity of \cref{eq:RandKick} may pose a stronger constraint \(\Delta v_\mathrm{drag}/\Delta v_{ij} < r_\mathrm{diss}^{-1}\) on the time step. Thus, one must choose \(\tau < r_\mathrm{diss}^{-1}\).  

If multiple elastic or dissipative processes contribute, summing the respective drag forces (\(F_\mathrm{drag}\)) and the energy losses (\(\Delta E\)) is straightforward. One can therefore define effective simulation parameters, 
\begin{equation}
    \sigma^\mathrm{eff}_\mathrm{T} = \sum\limits_r \sigma_\mathrm{T}^r, \quad r^\mathrm{eff}_\mathrm{diss} = \frac{1}{\sigma^\mathrm{eff}_\mathrm{T}}\sum\limits_r\sigma_\mathrm{T}^rr_\mathrm{diss}^r,
\end{equation}
which can be precomputed and used without further modification of the algorithm.

In summary, the dissipative extension of the fSIDM algorithm consists of a drag velocity enhanced by \(r_\mathrm{diss}\) (see Eq. \ref{eq:rdiss}), a time step reduced by \(r_\mathrm{diss}\), and a slightly modified expression for the stochastic velocity kick (see Eq. \ref{eq:RandKick}). Multiple interaction channels can be incorporated through effective parameters. 

Both the cross section (\(\sigma_\mathrm{T}/m_\chi\)) and the dissipation parameter (\(r_\mathrm{diss}\)) generally depend on the relative velocity. In this work, we limited ourselves to constant parameters, as a first step. This corresponds to the situation in which neither the cross section nor the dissipated energy introduce an additional velocity scale. The dissipated energy must thus be proportional to the kinetic energy. In this case, the local cooling rate is given by
\begin{equation}\label{eq:fSIDMcoolingRate}
    C(\rho,\nu) \equiv -\frac{\mathrm{d}E}{\mathrm{d}V\,\mathrm{d}t} = \frac{8}{\sqrt{\uppi}}\frac{\sigma_\mathrm{T}}{m_\chi}(r_\mathrm{diss}-1)\rho^2\nu^3
,\end{equation}
and the heat conduction rate is proportional to \(\sigma_\mathrm{T}/m_\chi r_\mathrm{diss}\), as shown in Appendix~\ref{appsec:coolingAndCondRate} by thermally averaging over scatterings of different relative velocities. In Appendix~\ref{sec:Validation} we cross-check our implementation against solutions of differential equations that can be derived with these expressions in the absence of gravity.

\subsection{Rarely self-interacting dark matter (rSIDM) with dissipation}\label{sec:dissipativerSIDM}
Cross sections that are not sufficiently dominated by forward scatterings to be modeled by the fSIDM algorithm must be explicitly sampled \citep{Burkert2000,Kochanek2000,Dave2001,Colin2002,Robertson2017} or split into small-angle and large-angle scatterings \citep{Arido2025}. In general, the scattering probability then follows from the differential cross section \(\frac{\mathrm{d}\sigma}{\mathrm{d}\Omega\,\mathrm{d}k^0}\) and depends not only on the scattering angle (\(\theta\)) and relative velocity (\(\Delta v_{ij}\)) but also on the radiated energy (\(k^0\)). The randomly sampled \(\theta\) and \(k^0\) would then be applied to the directions and magnitudes of the outgoing DM particles, respectively.

To compare with the dissipative fSIDM case, we implemented a simple example of this scheme for an isotropic cross section given by a constant \(\sigma^\mathrm{iso}_\mathrm{tot}/m_\chi\) and a constant dissipation parameter \(f_\mathrm{diss}\) that encodes the fraction of the CMS kinetic energy dissipated per interaction \citep[cf.][]{Shen2021}. The scattering probability of two particles \(i\) and \(j\) is then given by 
\begin{equation}
    P_{ij} = \frac{\sigma^\mathrm{iso}_\mathrm{tot}}{m_\chi}\Delta v_{ij} \frac{\Delta t}{m} \int\mathrm{d}V\,\rho_i\rho_j \ll 1
\end{equation}
with the time step ensuring its smallness. Since this is only computationally feasible if the scatterings are sufficiently rare, \citet{Kahlhoefer2014} named these models rSIDM (rarely self-interacting dark matter) as opposed to fSIDM. 

If two simulation particles happen to interact, they scatter into a random direction in the CMS frame, emerging back-to-back with speeds reduced to
\begin{equation}
    v' = v\sqrt{1 - f_\mathrm{diss}}.
\end{equation}
One can show that this generates a local cooling rate,
\begin{equation}\label{eq:rSIDMcoolingRate}
    C(\rho,\nu) = \frac{4}{\sqrt{\uppi}}\frac{\sigma^\mathrm{iso}_\mathrm{tot}}{m_\chi}f_\mathrm{diss} \rho^2 \nu^3,
\end{equation}
similar to \cref{eq:fSIDMcoolingRate}. The precise correspondence is established in \cref{sec:fSIDMrSIDMmatching}.

\section{Simulation of an isolated, dissipative SIDM halo}\label{sec:Results}
Having validated our code (introduced in \cref{sec:Method}) in controlled test setups without gravity (see~Appendix~\ref{sec:Validation}), we next applied it to an isolated DM halo (described in \cref{sec:Setup}). The goal was to identify qualitative changes in the gravothermal evolution with respect to the elastic SIDM case. We ran two series of simulations with increasing dissipation at constant heat conduction (\cref{sec:IncreasingDissipation}) and vice versa (\cref{sec:ConstantDissipation}) to disentangle their effects. Finally, we compared the halo evolutions simulated with fSIDM, i.e., with a forward-dominated cross section, with simulations employing an isotropic rSIDM cross section (\cref{sec:fSIDMrSIDMmatching}).

\subsection{Setup}\label{sec:Setup}
Using \texttt{SpherIC}\footnote{\url{https://bitbucket.org/migroch/spheric/src/main/},\\maintained version: \url{https://github.com/mtryan83/SpherIC}} \citep{GarrisonKimmel2013}, we generated a discrete realization of a Navarro-Frenk-White (NFW) halo \citep{Navarro1997} with scale radius \(r_\mathrm{s} = 3.6\,\mathrm{kpc}\) and characteristic density \(\rho_0 = 4\rho(r_\mathrm{s}) =  7.09\times 10^6\,\mathrm{M}_\odot\,\mathrm{kpc}^{-3}\), represented by \(N = 5\times 10^6\) particles. The code introduces an exponential cutoff chosen beyond \(r_\mathrm{cutoff} = 55\,\mathrm{kpc}\) to render the total halo mass finite at a value of \(M = 10^{10}\,\mathrm{M}_\odot\). We took this mass to equal \(M_{200\mathrm{c}}\), which corresponds to a radius \(r_{200\mathrm{c}} \simeq 44\,\mathrm{kpc}\), defined by an average enclosed density of 200 times the critical density of the Universe at \(z=0\). The virial radius thus lies well within the cutoff radius. The concentration parameter \(c_{200\mathrm{c}} \equiv r_{200\mathrm{c}}/r_\mathrm{s} \simeq 12\) is somewhat below the median mass-concentration relation, but is well within the reported scatter \citep{Dutton2014,Diemer2019}. 

Even at this comparatively high particle number, we find a percent-level scatter in the initial energy of the sampled halo relative to the analytic expectation. This variation translates into relative differences in the collapse time that are roughly ten times larger (see~Appendix~\ref{appsec:ConvergenceTest}) because halos of lower initial energy begin dissipation-driven collapse earlier. To mitigate this effect, we generated multiple realizations and selected one with a relative energy deviation of only \(3\times 10^{-5}\). 

Following \citet{Mace2024}, we further verified that, for this realization, the collapse time is insensitive to gravitational softening lengths in the range \(\epsilon = 4.0 \text{--} 63\,\mathrm{pc}\). We adopted \(\epsilon = 10\,\mathrm{pc}\) for all the simulations presented below.

We compared all \(N\)-body results with outputs from the fluid model code \textsc{GravothermalSIDM}\footnote{\url{https://github.com/kboddy/GravothermalSIDM}} \citep{Nishikawa2020,Outmezguine2023,GadNasr2024}, augmented by the cooling rate, \(C(\rho,\nu)\), from \cref{eq:fSIDMcoolingRate} as proposed by \citet{Essig2019}. Details are provided in Appendix~\ref{appsec:FluidModel}. In all subsequent figures, the corresponding fluid-model results are shown in the background for reference.

\begin{figure}
    \centering
    \includegraphics[width = \hsize]{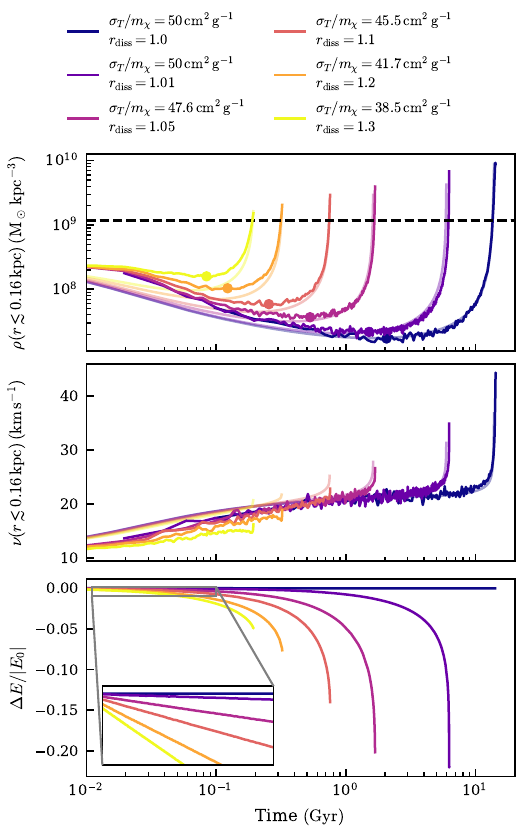}
    \caption{Time evolution of the central density (\textit{top}), velocity dispersion (\textit{middle}), and total dissipated energy (\textit{bottom}) of an isolated halo with different initial dissipation rates. Solid lines (foreground) show the \(N\)-body results, while faint lines (background) indicate the corresponding fluid-model predictions. The dashed horizontal line marks five times the initial central density, which is used to define the collapse time, \(t_\mathrm{coll}\). Colored dots indicate the minimum central density, which defines the core formation time, \(t_\mathrm{core}\).}
    \label{fig:TimeEvolutionIncreasingDissipation}
\end{figure}

\subsection{Varying dissipation at constant heat conduction}\label{sec:IncreasingDissipation}
We first considered simulations with different initial cooling rates \(\propto \sigma_\mathrm{T}/m_\chi(r_\mathrm{diss} - 1)\) at fixed initial conduction rate \(\propto \sigma_\mathrm{T}/m_\chi r_\mathrm{diss}\) (cf. Eq. \ref{eq:TwoComponentODE}). \Cref{fig:TimeEvolutionIncreasingDissipation} shows the time evolution of the central density and velocity dispersion (measured within \(r\lesssim 0.16\,\mathrm{kpc}\)), as well as the total dissipated energy. The inset in the bottom panel highlights the different initial cooling rates. 

The curves exhibit the characteristic gravothermal evolution of SIDM halos to varying degrees (e.g.,~\citealt{Tulin2018}, see the blue curve for elastic SIDM). During the initial core formation phase, the central density decreases before rising again during core collapse. We defined the collapse time, \(t_\mathrm{coll}\), through the condition\begin{equation}
    \rho(t_\mathrm{coll}, r\lesssim0.16\,\mathrm{kpc}) \equiv 5\rho(0, r\lesssim0.16\,\mathrm{kpc}),
\end{equation}
indicated by the dashed horizontal line. Across the range of dissipation parameters considered, \(t_\mathrm{coll}\) decreases by nearly two orders of magnitude, which is in qualitative agreement with \citet{Essig2019} and \citet{Huo2020}.

The minimum central density (after Gaussian filtering over five data points) is marked by the dots in the top panel and defines the core formation time, \(t_\mathrm{core}\). With increasing dissipation, the corresponding core density, \(\rho(t_\mathrm{core})\), increases, while the velocity dispersion, \(\nu(t_\mathrm{core})\), decreases. By the end of the simulation, the total dissipated energy is smaller for stronger dissipation, reflecting the shorter evolution time prior to collapse. 

The fluid model reproduces these trends qualitatively but predicts lower densities and higher velocity dispersions during core formation. Nevertheless, the collapse times agree well, within 5\,\%, even though the model's long-mean-free-path (LMFP) conductivity was calibrated solely on the elastic simulation (see~Appendix~\ref{appsec:FluidModel}).

\begin{figure}
    \centering
    \includegraphics[width = \hsize]{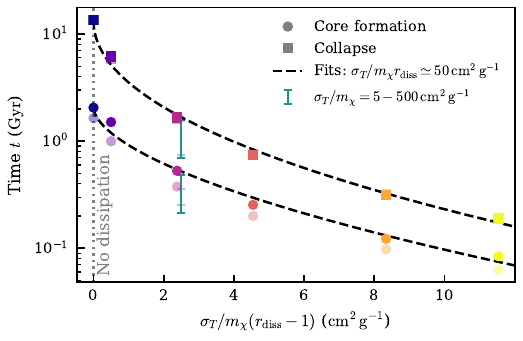}
    \caption{Core formation times (circles) and collapse times (squares) as a function of the dissipation strength \(\sigma_\mathrm{T}/m_\chi(r_\mathrm{diss}-1)\) at fixed conduction strength \(\sigma_\mathrm{T} r_\mathrm{diss} \simeq 50\,\mathrm{cm}^2\,\mathrm{g}^{-1} \). The colors are consistent with \cref{fig:TimeEvolutionIncreasingDissipation}. Green error bars indicate results obtained by varying \(\sigma_\mathrm{T}/m_\chi\) at fixed \(\sigma_\mathrm{T}/m_\chi(r_\mathrm{diss}-1) = 2.5\,\mathrm{cm}^2\,\mathrm{g}^{-1}\) (see~\cref{sec:ConstantDissipation}). Faint markers show the corresponding fluid-model results. Dashed lines indicate best-fit relations given by \cref{eq:FitIncreasingDissipation} with the parameters listed in \cref{tab:FitParameters}.} 
    \label{fig:IncreasingDissipationCorrelation}
\end{figure}

\begin{figure*}
    \centering
    \includegraphics[width = \hsize]{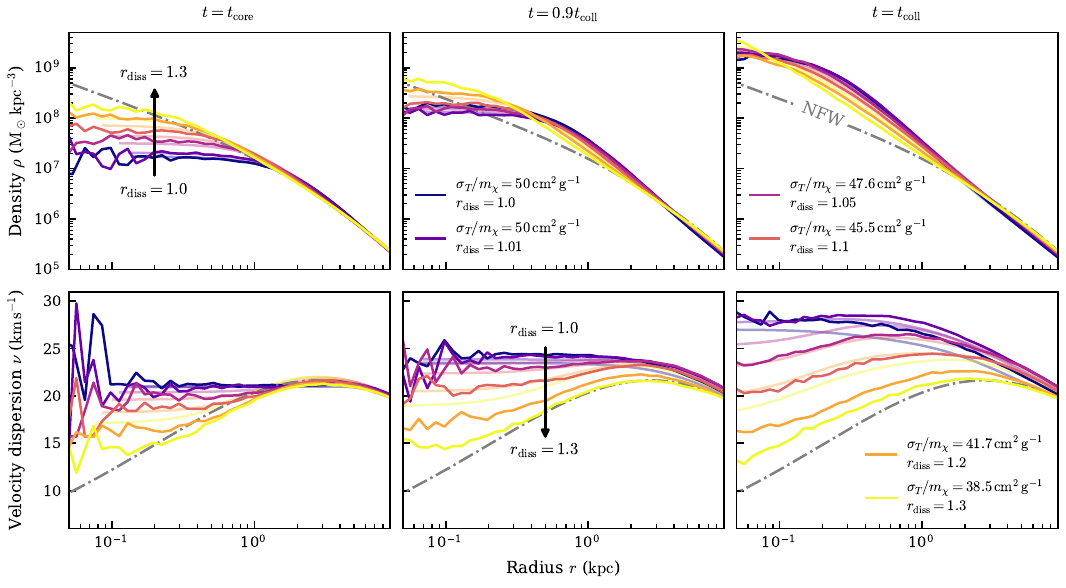}
    \caption{\textit{Top row:} Density profiles at \(t = t_\mathrm{core}\), \(t = 0.9t_\mathrm{coll}\), and \(t = t_\mathrm{coll}\) for different dissipation strengths. The initial NFW halo is shown in gray for reference, and corresponding fluid-model profiles are shown as faint lines. 
    \textit{Bottom row:} Corresponding velocity dispersion profiles.}
    \label{fig:IncreasingDissipationProfiles}
\end{figure*}

\Cref{fig:IncreasingDissipationCorrelation} shows the dependence of the core formation and collapse times on the parameter combination \(\sigma_\mathrm{T}/m_\chi(r_\mathrm{diss} - 1)\), to which the cooling rate, \(C(\rho,\nu)\), is proportional (see Eq. \ref{eq:fSIDMcoolingRate}). Both times decrease rapidly with increasing dissipation and are well fitted by a stretched exponential,
\begin{equation}\label{eq:FitIncreasingDissipation}
    t_\mathrm{core/coll}\left(x\equiv\frac{\sigma_\mathrm{T}}{m_\chi}(r_\mathrm{diss}-1)\right) = t_\mathrm{core/coll}(x=0)\exp\left[-\left(\frac{x}{x_0}\right)^\alpha\right]
,\end{equation}
where the two fit parameters are listed in \cref{tab:FitParameters} and \(t_\mathrm{core/coll}(x=0)\) are fixed to the respective non-dissipative values. Both increasing and decreasing the cross section relative to the fitted value of \(\sigma_\mathrm{T}/m_\chi r_\mathrm{diss} \simeq 50\,\mathrm{cm}^2\,\mathrm{g}^{-1}\) accelerate the evolution, as indicated by the green bars. This non-monotonic behavior is discussed in detail in \cref{sec:ConstantDissipation}. Generally, the impact of \(\sigma_\mathrm{T}/m_\chi\) is much smaller than that of \(r_\mathrm{diss}\). To what extent the fit parameters \(\alpha\) and \(x_0\) universally apply under the change of cross section and initial conditions beyond the rescaling symmetry introduced in Appendix~\ref{appsec:Rescaling} remains to be investigated. 

\begin{table}[b]
\caption{Fit parameters for \cref{eq:FitIncreasingDissipation}.}
\label{tab:FitParameters} 
\centering           
\begin{tabular}{c | c c}   
\hline\hline               
 & \(x_0\) (\(\mathrm{cm}^2\,\mathrm{g}^{-1}\)) & \(\alpha\) \\
\hline                     
\(t_\mathrm{coll}\) & \(0.54\pm0.10\) & \(0.48\pm0.03\)\\
\(t_\mathrm{core}\) & \(1.4\pm0.3\) & \(0.58\pm0.07\)\\
\hline                                 
\end{tabular}
\tablefoot{The first row refers to the fit of the collapse time, the second row to that of the core formation time. \(x_0\) normalizes the argument of the stretched exponential with exponent \(\alpha\).}
\end{table}

\Cref{fig:IncreasingDissipationProfiles} shows the corresponding density and velocity dispersion profiles at \(t = t_\mathrm{core}\), \(t = 0.9t_\mathrm{coll}\), \(t = t_\mathrm{coll}\). With increasing dissipation, the resulting core becomes denser, less extended, and further from isothermality. While both the density and velocity dispersion increase toward collapse, the core properties described above persist, with the velocity dispersion gradient remaining positive at small radii. At \(t_\mathrm{coll}\) -- shortly before the \(N\)-body simulation breaks down because time steps become prohibitively small -- the density profile has a less pronounced core and exhibits a shallower outer slope  for strong dissipation. We discuss the underlying reason for this behavior in \cref{sec:Discussion}. The fluid model can capture these features qualitatively but systematically overestimates the velocity dispersion at larger radii as well.

\subsection{Varying heat conduction at constant dissipation}\label{sec:ConstantDissipation}
\begin{figure}
    \centering
    \includegraphics[width = \hsize]{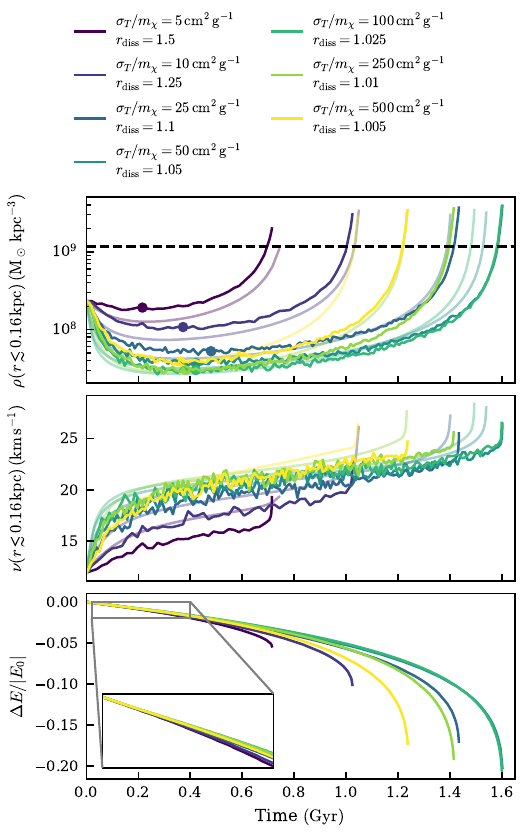}
    \caption{Time evolution of the central density (\textit{top}), velocity dispersion (\textit{middle}), and total dissipated energy (\textit{bottom}) as in \cref{fig:TimeEvolutionIncreasingDissipation}, but for a set of simulations of an isolated halo with dissipative fSIDM parameters chosen to correspond to equal initial cooling rates (see the inset).}
    \label{fig:TimeEvolutionConstantDissipation}
\end{figure}

\begin{figure}
    \centering
    \includegraphics[width = \hsize]{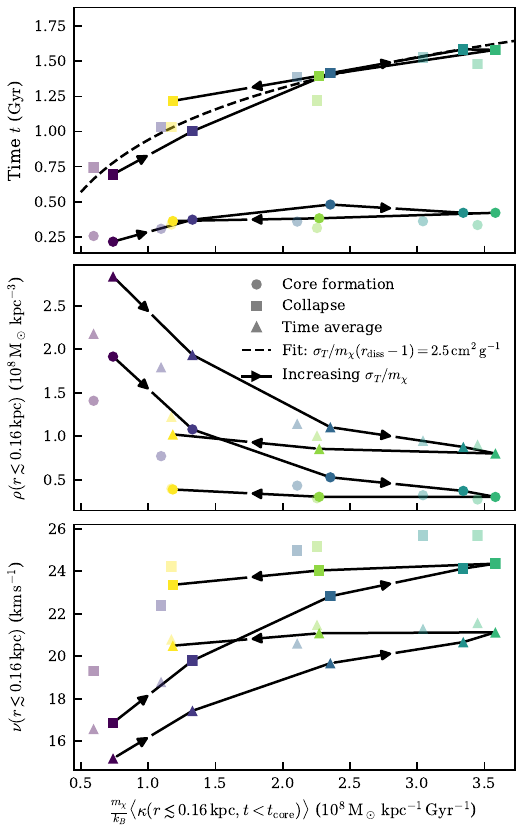}
    \caption{Correlation between the time-averaged heat conductivity \(\langle\kappa(t<t_\mathrm{core})\rangle\) and the core formation and collapse time (\textit{top}), the minimum and time-averaged central density (\textit{middle}), and the maximum and time-averaged central velocity dispersion (\textit{bottom}). Colors are consistent with \cref{fig:TimeEvolutionConstantDissipation}, and arrows indicate an increasing cross section. We omit \(\rho(t_\mathrm{coll})\) as it is constant by definition and \(\nu(t_\mathrm{core})\) as it is very similar to \(\langle\nu(t<t_\mathrm{coll})\rangle\) but noisier.}
    \label{fig:ConstantDissipationCorrelation}
\end{figure}

\begin{figure*}
    \centering
    \includegraphics[width = \hsize]{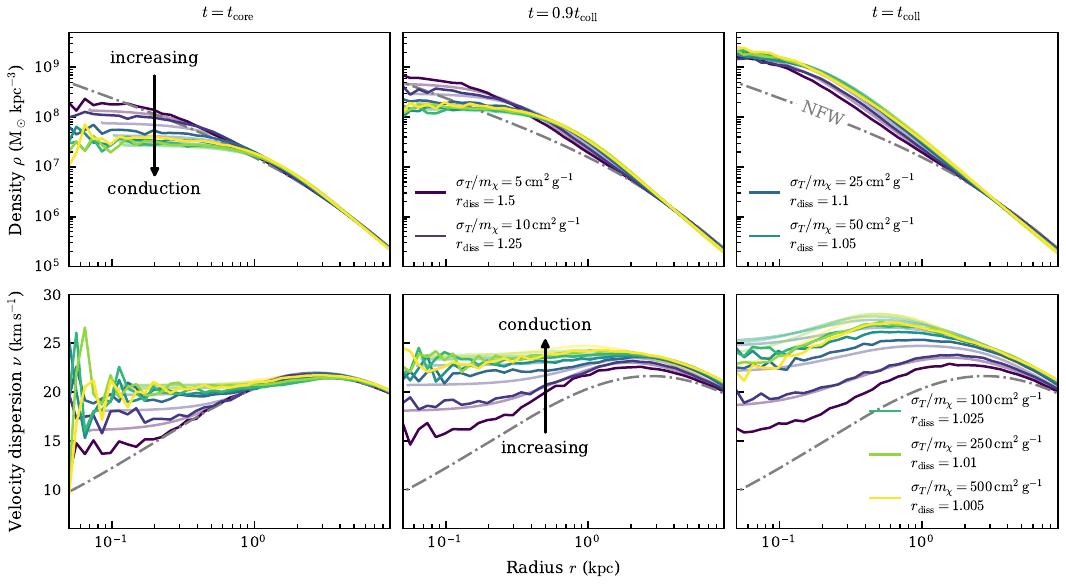}
    \caption{Density (\textit{top row}) and velocity dispersion (\textit{bottom row})  profiles as in \cref{fig:IncreasingDissipationProfiles} but varying the cross section at a fixed initial cooling rate. Some trends reverse with increasing cross section, which we attribute to the non-monotonic behavior of the heat conduction.}
    \label{fig:ConstantDissipationProfiles}
\end{figure*}

As a second set of simulations, we varied the transfer cross section in the range \(5\,\mathrm{cm}^2\,\mathrm{g}^{-1} < \sigma_\mathrm{T}/m_\chi < 500\,\mathrm{cm}^2\,\mathrm{g}^{-1}\) while keeping the initial cooling rate \(C(\rho,\nu) \propto \sigma_\mathrm{T}/m_\chi(r_\mathrm{diss}-1)\) fixed. \Cref{fig:TimeEvolutionConstantDissipation} shows the time evolution of the central density and velocity dispersion analogous to \cref{fig:TimeEvolutionIncreasingDissipation}. In the bottom panel, all initial slopes of the energy loss are indeed identical, before the halos diverge onto different evolutionary tracks. 

The variation in collapse and core formation times is much smaller than in the previous case, amounting to less than a factor of three. Both increase with the cross section for \(\sigma_\mathrm{T}/m_\chi \lesssim 50\,\mathrm{cm}^2\,\mathrm{g}^{-1}\), before decreasing at larger cross sections. The central density and velocity dispersion exhibit a similar non-monotonic trend. It is driven by the heat conductivity, \(\kappa\), which increases at small cross sections, where it is limited by the self-interaction strength \(\propto \sigma_\mathrm{T}\) (LMFP regime). At large cross sections, the decreasing mean free path \(\propto\sigma_\mathrm{T}^{-1}\) suppresses conduction, causing \(\kappa\) to decline; this is the  short-mean-free-path (SMFP) regime.

\Cref{fig:ConstantDissipationCorrelation} shows the correlation between quantities extracted from the halo evolution and the central fluid-model heat conductivity, time-averaged up to \(t_\mathrm{core}\). Among several tested definitions -- \(\kappa(t=0)\), \(\kappa(t_\mathrm{core})\), \(\kappa(t_\mathrm{coll})\), \(\langle\kappa(t < t_\mathrm{coll})\rangle\), and \(\langle\kappa(t < t_\mathrm{core})\rangle\) -- the last of these shown in the figure yields the tightest correlation. 

An empirical relation of the form
\begin{equation}\label{eq:FitConstantDissipation}
\begin{split}
    t_\mathrm{coll}(\langle\kappa\rangle) &= (0.54\pm 0.07)\,\mathrm{Gyr}\,\log\left(\frac{\langle\kappa(t<t_\mathrm{core})\rangle}{10^8 \,\mathrm{M}_\odot\,\mathrm{kpc}^{-1}\,\mathrm{Gyr}^{-1}}\right)\\ 
    &+ (0.94\pm 0.06)\,\mathrm{Gyr}
\end{split}
\end{equation}
provides a good description of the collapse time over the explored range, though a residual dependence on the cross section remains, as indicated by the arrows in \cref{fig:ConstantDissipationCorrelation}. The fit systematically overestimates \(t_\mathrm{coll}\) at small cross sections and underestimates it at large cross sections. For the central density and velocity dispersion, the correlations are weaker. We therefore did not fit them.

The identified correlations clearly indicate that heat conduction is the physical mechanism underlying the non-monotonic behavior with varying cross section. This interpretation is further supported by the velocity dispersion profiles in \cref{fig:ConstantDissipationProfiles}, which first become more isothermal with increasing cross section, indicating stronger heat conduction, but then slightly less isothermal again at large cross sections. Correspondingly, the core becomes less dense and more extended when heat conduction is most efficient. In this sense, conduction counteracts the effects of dissipation identified in \cref{fig:IncreasingDissipationProfiles} and thus delays collapse. Through this mechanism, the non-monotonic dependence of the heat conductivity on the cross section at the transition from the LMFP to the SMFP regime can propagate to the core formation and collapse time and the central density and velocity dispersion.

The predictions of the gravothermal fluid model are significantly less accurate than in \cref{sec:IncreasingDissipation}. For the smallest cross section, the collapse time is overestimated by 7\,\%, while for the largest cross section, it is underestimated by 16\,\%. This could be explained by an effective decrease in the calibration parameter \(\beta\) (see~Appendix~\ref{appsec:FluidModel}) governing the LMFP heat conduction in the fluid model. Qualitatively, \citet{Mace2026} found the same behavior for elastic SIDM. They argue that an alternative interpolation rule between the LMFP and SMFP regimes can absorb this dependence while keeping \(\beta\) constant. As previously, the fluid model underestimates the density after core formation and overestimates the velocity dispersion.

\subsection{Matching condition between an isotropic (rSIDM) and forward-dominated (fSIDM) cross section}\label{sec:fSIDMrSIDMmatching}
We ran a set of elastic and dissipative rSIDM simulations (see \cref{sec:dissipativerSIDM}) with different isotropic cross sections, \(\sigma_\mathrm{tot}^\mathrm{iso}\). Under certain matching conditions to the fSIDM parameters, we find a very similar time evolution of isolated halos. In the elastic case, the collapse times obtained using rSIDM and fSIDM agree within 1.1\,\% for \(\sigma_\mathrm{tot}^\mathrm{iso} = 50\,\mathrm{cm}^2\,\mathrm{g}^{-1}\) and \(\sigma_\mathrm{T}^\mathrm{fSIDM} = 150\,\mathrm{cm}^2\,\mathrm{g}^{-1}\) (see Fig. \ref{fig:Matching}, top panel). The difference by a factor of three is expected since SIDM halos of equal viscosity cross section,
\begin{equation}
    \sigma_\mathrm{V} = 4\uppi\int\limits_0^1\mathrm{d}\!\cos\theta (1-\cos^2\theta)\frac{\mathrm{d}\sigma}{\mathrm{d}\Omega} = \begin{cases}
        2\sigma_\mathrm{tot}^\mathrm{iso}/3,&\text{rSIDM}\\ 2\sigma_\mathrm{T}^\mathrm{fSIDM},&\text{fSIDM}
    \end{cases}
,\end{equation}
are known to evolve similarly \citep{Yang2022,Sabarish2024}.\footnote{Note that there are conventions for the viscosity cross section in the literature that differ by numerical factors. The relation between \(\sigma_\mathrm{tot}^\mathrm{iso}\) and \(\sigma_\mathrm{T}^\mathrm{fSIDM}\) is independent of the precise choice.}

With dissipation, we expect a similar time evolution when the cooling rates given in \cref{eq:fSIDMcoolingRate,eq:rSIDMcoolingRate} are equal. This implies the full matching condition
\begin{equation}\label{eq:fSIDMrSIDMmatching}
    f_\mathrm{diss} \leftrightarrow \frac{2}{3}(r_\mathrm{diss} - 1),\quad \sigma_\mathrm{tot}^\mathrm{iso} \leftrightarrow 3\sigma_\mathrm{T}^\mathrm{fSIDM}.
\end{equation}
For the representative example in the bottom panel of Fig. \ref{fig:Matching}, it is confirmed with a relative difference of 2.5\,\% in the collapse times (solid blue curve). Choosing a larger \(f_\mathrm{diss}\) at the same cross section significantly accelerates the collapse (dashed blue curve). A simultaneous increase in \(f_\mathrm{diss}\) and decrease in \(\sigma_\mathrm{tot}^\mathrm{iso}\) at a constant cooling rate has a much smaller impact (orange curve). Yet, it still decreases the collapse time due to the weaker heat conduction, which enables a more efficient central dissipation (cf.~\cref{sec:ConstantDissipation}).

The existence of such simple matching conditions demonstrates that the angular dependence of the cross section plays a minor role in the evolution of isolated halos. It may be more relevant for satellites and mergers, as simulations of elastic SIDM suggest \citep[e.g.,][]{Fischer2021,Klemmer2026}.

\section{Discussion}\label{sec:Discussion}
In this section, we first provide a physical intuition for the simulation results found, employing the fluid model to discuss the radially resolved halo energy budget (\cref{sec:energyBudget}). Then, we draw the connection to astrophysical observables by projecting along the line of sight (\cref{sec:ProjectedQuantities}) and discuss whether the projected density profile of the exotic compact object in JVAS~B1938+666 inferred by \citet{Vegetti2026} could be explained by dissipative SIDM (\cref{sec:VegettiObservation}).

\subsection{Gravothermal evolution with dissipation}\label{sec:energyBudget}
\begin{figure*}
    \centering
    \includegraphics[width = \hsize]{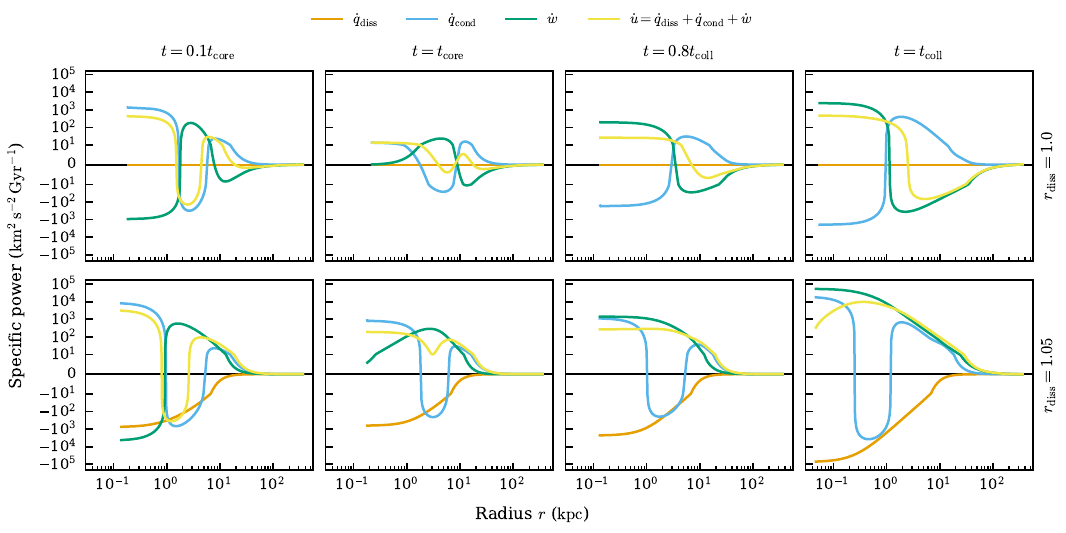}
    \caption{Specific power, i.e., energy change per mass bin, due to dissipation (\(\dot{q}_\mathrm{diss}\)), conduction (\(\dot{q}_\mathrm{cond}\)), and mechanical work (\(\dot{w}\)), together with their sum (i.e., the specific internal energy change, \(\dot{u}\)) as a function of radius. The columns show different times relative to the core formation time (\(t_\mathrm{core}\)) and collapse time (\(t_\mathrm{coll}\)). The top row was obtained from an elastic fluid model run with \(\sigma_\mathrm{T}/m_\chi = 50\,\mathrm{cm}^2\,\mathrm{g}^{-1}\), the bottom row from a dissipative run with \(r_\mathrm{diss} = 1.05\) and the same cross section.}
    \label{fig:EnergyBudget}
\end{figure*}
To interpret the results presented in \cref{sec:Results}, we adopted the simplified picture of the gravothermal fluid model (see~Appendix~\ref{appsec:FluidModel}), which provides a qualitatively accurate description as demonstrated. In this picture, heat conduction proceeds from regions of higher to lower velocity dispersion, i.e., both inward and outward from the maximum of the initial NFW profile. As heat is transported into the center, the central velocity dispersion increases, causing particles to migrate to larger orbits. This leads to an outward mass flow, reducing the central density and thereby forming a core.

The structure of the core depends on the competition between conductive heating and dissipative cooling, the latter being most efficient at high densities. If conduction dominates, the core is heated and expands, resulting in a lower central density. Conversely, if cooling dominates, the core remains colder and denser. Crucially, a fully isothermal core can only form in the absence of dissipation.

In the presence of dissipation, we find that a more or less pronounced off-center (\(r\neq 0\)) maximum in the velocity dispersion profile persists throughout the evolution (see~\cref{fig:IncreasingDissipationProfiles,fig:ConstantDissipationProfiles}, bottom rows). As a result, the direction of heat conduction in the inner region does not reverse, in contrast to the elastic case. The core loses energy locally through dissipation rather than conduction, without heating outer regions of the halo. Inward heat conduction partially compensates for the energy loss, thereby delaying collapse. This explains why stronger conduction leads to longer collapse times in our simulations (see~\cref{fig:TimeEvolutionConstantDissipation}), opposite to the trend observed in the elastic case.

Using the fluid model, we can explicitly visualize this by decomposing the rate of specific energy change into contributions from conduction, \(\dot{q}_\mathrm{cond}\), dissipation, \(\dot{q}_\mathrm{diss}\), and gravitational work, \(\dot{w}\) as a function of radius and time (see~\cref{fig:EnergyBudget}). The dissipation term (\(\dot{q}_\mathrm{diss}\)) is nonpositive at all radii and peaks in magnitude at small radii where the density is highest. As expected, it vanishes in the elastic case of \(r_\mathrm{diss} = 1\) (top row). 

In contrast, conduction (\(\dot{q}_\mathrm{cond}\)) only redistributes energy within the halo and therefore integrates to zero over mass. Positive work (\(\dot{w}>0\)) is performed on regions that are compressed, while expanding regions perform negative work (\(\dot{w}<0\)). By construction of the fluid model, the sum of all contributions yields the change in specific internal energy, which determines the velocity dispersion.

Initially, regions near the maximum of the velocity dispersion profile at \(r\sim r_\mathrm{s} = 3.6\,\mathrm{kpc}\) lose energy via conduction, transferring heat to both the core and the outer halo. The work term (\(\dot{w}\)) reflects the gravitational response: heated regions generally expand, while cooling regions contract. These contributions do not exactly balance, resulting in a net change in internal energy and hence velocity dispersion.

In the elastic case (top row), this process gradually erases the maximum in the velocity dispersion profile. Once the maximum is at the center, the direction of heat conduction reverses (see\ top-right panels), and the collapse phase begins. The negative heat capacity of self-gravitating systems manifests itself in the opposite signs of \(\dot{q}_\mathrm{cond}\) and \(\dot{u}\). Although energy is transported outward, the core heats up due to the strong compressional work (\(\dot{w}>0\)), which dominates over conductive cooling.

In the dissipative case (bottom row), the early evolution (\(t = 0.1t_\mathrm{core}\)) closely resembles the elastic scenario. However, at later times, dissipative cooling dominates over conduction. The halo responds with a global contraction (\(\dot{w} > 0\)), sufficiently strong to drive an increase in internal energy (\(\dot{u}>0\)). This occurs even before the maximum in the velocity dispersion is erased, distinguishing the dissipative from the elastic collapse pathway. 

As a consequence, the direction of heat conduction does not reverse at any stage of the evolution. Instead of conduction-driven collapse, the halo contracts due to continuous energy loss. This overall contraction enhances the central mass concentration and may seed the formation of compact objects, as we discuss in the following section.

\subsection{Consequences for lensing observables}\label{sec:ProjectedQuantities}
\begin{figure*}
    \centering
    \includegraphics[width = \hsize]{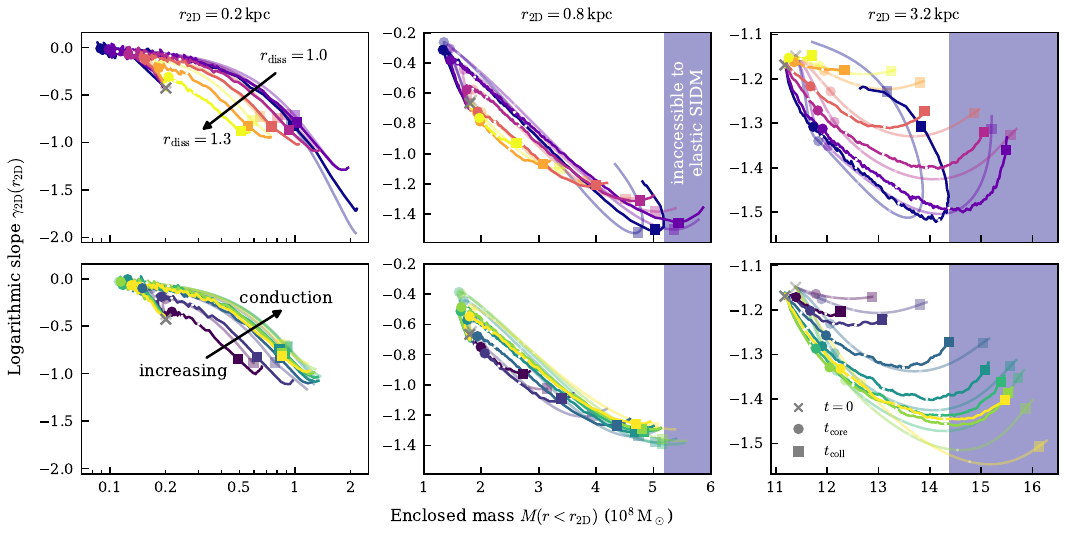}
    \caption{Evolutionary tracks of all simulations in the plane of projected enclosed mass and the logarithmic slope of the surface density profile. \textit{Top row}: Simulations with varying dissipation (\cref{sec:IncreasingDissipation}). \textit{Bottom row}: Simulations with varying heat conduction (\cref{sec:ConstantDissipation}). The colors have the same meanings as in \cref{fig:TimeEvolutionIncreasingDissipation,fig:TimeEvolutionConstantDissipation}. Faint background lines represent the fluid model, and white dots mark time intervals of \(0.2t_\mathrm{coll}\).} 
    \label{fig:2DevolutionTracks}
\end{figure*}
To connect our simulation results to observables accessible to strong-lensing surveys, we considered the time evolution of quantities derived from the projected surface density, \(\Sigma(r_\mathrm{2D})\), (see Appendix~\ref{appsec:ProjectedQuantities} for definitions and calculation details). In particular, \cref{fig:2DevolutionTracks} shows the evolutionary tracks of all simulations in the plane spanned by the projected enclosed mass, \(M(r<r_\mathrm{2D})\), and the logarithmic slope of the surface density profile, \(\gamma_\mathrm{2D}(r_\mathrm{2D})\), evaluated at \(r = 0.2\,\mathrm{kpc}\), \(r = 0.8\,\mathrm{kpc}\), and \(r = 3.2\,\mathrm{kpc}\) \citep[cf.][]{Fischer2025a}. The arrows indicate increasing dissipation (top row) and heat conduction (bottom row). 

At \(r = 0.2\,\mathrm{kpc}\), the core formation and collapse phase are clearly distinguishable. Initially, the enclosed mass decreases and the density profile becomes shallower until \(t_\mathrm{core}\) (marked by the circle), before both reverse. The overall variation of both quantities throughout the evolution generally decreases as dissipation dominates more strongly over heat conduction. This reflects the increasingly suppressed core formation. At \(r = 0.8\,\mathrm{kpc}\), the two phases remain visible for weak dissipation. For strong dissipation, however, both the enclosed mass and profile steepness increase monotonically from the beginning, indicating that the core radius is smaller than \(0.8\,\mathrm{kpc}\). At \(r = 3.2\,\mathrm{kpc}\), close to the scale radius, the enclosed mass and profile steepness increase immediately in all simulations, indicating that no core larger than this radius forms. 

Toward collapse, the non-dissipative simulation (blue, top row) exhibits a clear turnaround in both quantities for \(r \gtrsim 0.8\,\mathrm{kpc}\). The earlier one in \(\gamma_\mathrm{2D}\) reflects the formation of an indentation in the density profile caused by mass flowing both inward and outward from the radius where \(\dot{q}_\mathrm{cond} = 0\) (see \cref{fig:EnergyBudget}, top row). As this radius moves inward over time, outward mass transport eventually dominates, leading to the second turnaround in the enclosed mass. 

Dissipation largely suppresses this outward mass transport, at least at the radii considered here. As a result, the enclosed mass at \(t_\mathrm{coll}\) (squares) can exceed that of the elastic case and the region of parameter space shaded in blue in \cref{fig:2DevolutionTracks} becomes accessible for moderate dissipation. For very strong dissipation, however, the collapse time becomes so short that the enclosed mass at \(t_\mathrm{coll}\) is smaller, simply because the system has less time to evolve. At fixed absolute time, the intuitive trend is restored: stronger dissipation leads to larger enclosed mass, as particles lose energy and fall inward. The turnaround in the slope persists, but becomes progressively less pronounced with increasing dissipation relative to conduction.

Once again, the gravothermal fluid model can reproduce the physical trends described above. The precise values deviate slightly.  

\subsection{Reproducing observed lensing results}\label{sec:VegettiObservation}

We finally examined whether the gravothermal evolution of our simulated halos can reproduce the mass distribution inferred by \citet{Vegetti2026} for the compact object in JVAS~B1938+666 detected through gravitational imaging \citep{Powell2025}. Their analysis considered a broad class of density profiles, which agree best in enclosed projected masses at the two radii \(r = 20\,\mathrm{pc}\) and \(r = 90\,\mathrm{pc}\). We therefore adopted the corresponding best-fit values \(M(r < 20\,\mathrm{pc}) = (4.25\pm0.21)\times 10^5\,\mathrm{M}_\odot\) and \(M(r < 90\,\mathrm{pc}) = (1.167\pm0.039)\times 10^6\,\mathrm{M}_\odot\). Taking these two data points as given, we investigated whether our SIDM halos evolved with and without dissipation could match them after appropriate rescaling (see~Appendix~\ref{appsec:Rescaling}). Recall that they are initially described by an NFW profile and evolved in isolation. We further required an evolution time consistent with the observed redshift, \(z_\mathrm{obs}=0.881\). 

\begin{figure*}[tb]
    \centering
    \includegraphics[width = \hsize]{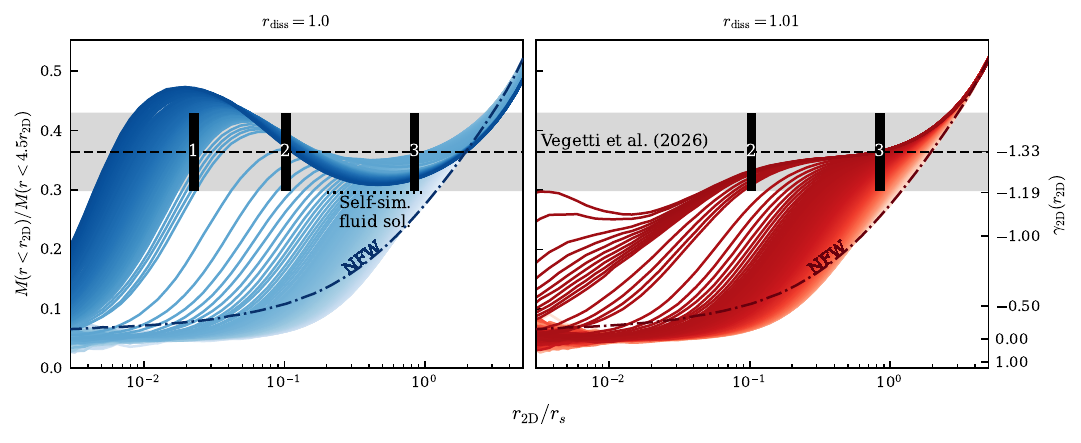}
    \caption{Ratio of projected enclosed masses \(M(r<r_\mathrm{2D})/M(r<4.5r_\mathrm{2D})\) as a function of the normalized inner radius (\(r_\mathrm{2D}/r_\mathrm{s}\)) for our elastic (\textit{left}) and least dissipative (\textit{right}) simulation. Darker colors correspond to later snapshots. The dashed horizontal line indicates the mass ratio of the exotic compact object in JVAS~B1938+666 inferred by \citet{Vegetti2026}, with the shaded region showing the \(3\sigma\) confidence interval. The dotted line represents the self-similar fluid model solution found by \citet{Balberg2002},  which the minimum appears to approach in the elastic case. We label three values of \(r_\mathrm{2D}/r_\mathrm{s}\), with some snapshots closely reproducing the observed mass ratio, which we consider as benchmarks (see \cref{table:RescaledQuantities}).}
    \label{fig:VegettiRatio}
\end{figure*}

In \cref{fig:VegettiRatio}, we first consider the corresponding dimensionless mass ratio \(M(r<r_\mathrm{2D})/M(r<4.5r_\mathrm{2D})\) as a function of the inner radius normalized by the NFW scale radius, \(r_\mathrm{2D}/r_\mathrm{s}\), for all snapshots of the elastic and least dissipative simulation (\(r_\mathrm{diss}=1.01\)). The gray band indicates the \(3\sigma\) confidence region around the observed mass ratio \(M(r<20\,\mathrm{pc})/M(r<90\,\mathrm{pc})=0.364\pm0.022\) (dashed line). Close to core collapse, both the elastic and dissipative SIDM halo match it over a wide range of scales.

Assuming that the density profile locally follows a simple power law, we can map the shown mass ratio to the slope of the projected density profile (\(\gamma_\mathrm{2D}\)) introduced in \cref{sec:ProjectedQuantities} (see the secondary axis). Except for the shrinking core, the late-time density profile of the collapsing halo is much steeper than the inner slope of the initial NFW profile and therefore in much better agreement with the observation in both cases. For \(r_\mathrm{diss} = 1.0\), the steep edge of the core causes an even larger mass ratio than required for the observation. It shifts inward and amplifies with time. The local minimum at \(r_\mathrm{2D}/r_\mathrm{s}\sim 0.5\) approaches the asymptotic slope \(\gamma_\mathrm{2D} = -1.19\) of the self-similar fluid-model solution in the LMFP limit \citep{Balberg2002}. For even larger radii, the curves resemble the initial NFW profile as self-interactions become inefficient. All this behavior generically falls within the confidence interval of the observed mass ratio over about two orders of magnitude in \(r/r_\mathrm{s}\).

For \(r_\mathrm{diss} = 1.01\), the maximum in the mass ratio disappears together with the steep section in the density profile (see~\cref{fig:IncreasingDissipationProfiles}). The curves suggest the existence of a similar attractor slope as in the elastic case, potentially even closer to the observed value. Our simulations with \(r_\mathrm{diss} > 1.01\) do not reproduce the observation within the time accessible to the simulation. Whether they would eventually do so if the simulations could be continued beyond numerical breakdown remains an open question and is a key limitation of the present approach.

\begin{table*}[bt]
\caption{Properties of the simulated halos, rescaled to match the size and mass of the profile from \citet{Vegetti2026}.}         
\label{table:RescaledQuantities} 
\centering
\resizebox{\hsize}{!}{
\begin{tabular}{c c | c c c c c c c c c}   
\hline\hline               
\# & \(r_\mathrm{diss}\) & \(r_\mathrm{s}\) (\(\mathrm{kpc}\)) & \(\rho_0\) (\(\mathrm{M}_\odot\,\mathrm{pc}^{-3}\)) & \(\sigma_\mathrm{T}/m_\chi\) (\(10^2\,\mathrm{cm}^2\,\mathrm{g}^{-1}\)) & \(t_\mathrm{evo}\) (\(\mathrm{Gyr}\))& \(\Delta t\) (\(\mathrm{Gyr}\)) & \(z_\mathrm{ini}\) & \(M_{200\mathrm{c}}\) (\(10^6\,\mathrm{M}_\odot\)) & \(c_{200\mathrm{c}}\)  & \(c_{200\mathrm{c}}^\mathrm{Diemer}(M_{200\mathrm{c}}, z_\mathrm{ini})\)\\    
\hline  

\textbf{1} & 1.0 & 0.88 & 0.0019 & 7.8 & 28 & 0.18 & \(\infty\) & \(-\) & \(-\) & \(-\) \\
\hline

\multirow{2}{*}{\textbf{2}} & 1.0 & \multirow{2}{*}{0.19} & \(0.063\) & \(1.0\) & \(4.8\) & 0.15 & \(4.0\) & \(6.3\) & \(5.9\) & \(4.9\)\\

& 1.01 & & \(0.054\) & \(1.2\) & \(2.3\) & 0.0037 & \(1.6\) & \(8.3\) & \(12\) & \(9.8\)\\
\hline

\multirow{2}{*}{\textbf{3}} & 1.0 & \multirow{2}{*}{0.024} & \(7.7\) & \(0.070\) & \(0.41\) & 0.24 & \(0.98\) & \(4.7\) & \(1.0\times10^2\) & \(13\)\\

& 1.01 & & \(7.1\) & \(0.076\) & \(0.20\) & 0.093 & \(0.93\) & \(4.3\) & \(1.0\times10^2\) & \(14\)\\
\hline                                 
\end{tabular}
}
\tablefoot{\# refers to the markers in \cref{fig:VegettiRatio}, \(r_\mathrm{diss}\) to the respective panel. \(r_\mathrm{s}\), \(\rho_0\), \(M_{200\mathrm{c}}\), and \(c_{200\mathrm{c}}\) are parameters of the initial NFW halo at redshift \(z_\mathrm{ini}\). \(\sigma_\mathrm{T}/m_\chi\) and \(t_\mathrm{evo}\) denote the rescaled cross section and evolution time required to explain the observation of the compact object by \citet{Vegetti2026}. \(\Delta t\) denotes the time the halo spends within the confidence interval during the simulated evolution. We compare the concentrations to the mass-concentration relation, \(c_{200\mathrm{c}}^\mathrm{Diemer}(M_{200\mathrm{c}},z_\mathrm{ini})\), from \citet{Diemer2019}. All quantities are rounded to two significant digits.}  
\end{table*}

Having considered only scale-invariant quantities until now, we explicitly rescaled the simulated halo to the size of the observed one in the following. We verified that \cref{fig:VegettiRatio} is fully invariant under the rescaling symmetry introduced in Appendix~\ref{appsec:Rescaling}, as expected. However, the mass ratio can differ significantly for initial conditions and cross sections that are not related by this symmetry. We highlight three representative values of \(r_\mathrm{2D}/r_\mathrm{s}\) for which the observed ratio can be reproduced within \(3\sigma\). In each case, we computed the corresponding rescaling parameters as described in Appendix~\ref{appsec:Rescaling} and list the rescaled physical halo properties in \cref{table:RescaledQuantities}.

The larger the radius, \(r_\mathrm{2D}/r_\mathrm{s}\), at which the halo shows the required mass ratio, the smaller the rescaled halo in extent. Since the required mass rescaling varies much less, a smaller halo implies a higher density (\(\rho_0\)), shorter  evolution time (\(t_\mathrm{evo}\)), and smaller cross section (\(\sigma_\mathrm{T}/m_\chi\)). Multiple snapshots falling within the confidence interval translate to a time interval (\(\Delta t\)), during which the observation can be explained. This interval is significantly shorter with dissipation. However, the values are only lower bounds, as the halo likely remains within the confidence interval beyond the simulation's breakdown. 

Using the observed lens redshift, \(z_\mathrm{obs}=0.881\), corresponding to a cosmic age of \(t(z_\mathrm{obs})=6.37\,\mathrm{Gyr}\) for the cosmology of \citet{PlanckCollaboration2020}, we inferred the redshift \(z_\mathrm{ini}\equiv z(t(z_\mathrm{obs})-t_\mathrm{evo})\) at which the halo was described by the NFW profile. With it, we computed the associated halo masses, \(M_{200\mathrm{c}}\), and concentrations, \(c_{200\mathrm{c}}\).\footnote{This approach is different from \citet{Vegetti2026}, who always use \(z_\mathrm{obs} = 0.881\) for the conversion. They further only consider a single snapshot and the data point at \(90\,\mathrm{pc}\).}

The largest and least dense halos (1) require evolution times exceeding the age of the Universe at \(z_\mathrm{obs}\) despite their large self-interaction cross sections. Neglecting potential accelerated evolution due to tidal stripping \citep{Nishikawa2020}, such systems are therefore disfavored. They would require even larger cross sections to collapse in time (recall also \(\sigma^\mathrm{iso}_\mathrm{tot} \leftrightarrow 3\sigma_\mathrm{T}\)). At the opposite extreme, the smallest and densest halos (3) undergo gravothermal collapse extremely rapidly even for very small cross sections, since \(t_\mathrm{coll} \propto \rho_0^{-3/2}\) \citep{Essig2019}. However, these halos require unrealistically large initial concentrations exceeding the \citet{Diemer2019} mass-concentration relation by more than \(5\sigma\).

The most plausible candidates are intermediate systems (2). These halos require concentrations less than \(1\sigma\) above the median relation, implying that a small but non-negligible fraction of halos in this mass range could collapse by \(z_\mathrm{obs}\). For comparable cross sections, dissipative simulations evolve roughly twice as fast as elastic simulations, demonstrating that even weak dissipation can significantly accelerate gravothermal evolution. Equivalently, dissipative halos could reproduce the observations with smaller self-interaction cross sections within the same evolution time.

Overall, our analysis shows that both elastic and weakly dissipative SIDM halos with cross sections \(\sigma_\mathrm{T}/m_\chi\sim 10^2\,\mathrm{cm}^2\,\mathrm{g}^{-1}\) -- which are largely unconstrained at masses as low as \(M_{200\mathrm{c}}\lesssim 10^7\,\mathrm{M}_\odot\) -- can naturally reproduce the compact mass distribution inferred by \citet{Vegetti2026} through the gravothermal evolution of moderately overdense isolated NFW halos within the available cosmic time. Dissipation further reduces the required cross section by accelerating the collapse process.

\section{Conclusions}
In this study we have introduced dissipation into the \(N\)-body formalism for frequent small-angle self-interactions, described by a drag force and transverse momentum diffusion. Compared to the elastic case, the drag is increased by a generally velocity-dependent dissipation parameter, \(r_\mathrm{diss}\). The associated energy loss is not fully restored by the transverse diffusion. Instead, it is chosen to reproduce the correct energy loss per collision of the underlying model. 

We applied the algorithm to an isolated DM halo with different velocity-independent transfer cross sections, \(\sigma_\mathrm{T}/m_\chi\), and dissipation parameters, \(r_\mathrm{diss}\). At constant heat conduction \(\propto \sigma_\mathrm{T}/m_\chi r_\mathrm{diss}\), cooling \(\propto \sigma_\mathrm{T}/m_\chi (r_\mathrm{diss}-1)\) can accelerate gravothermal evolution by orders of magnitude. Conversely, stronger heat conduction slightly inhibits collapse -- in contrast to the elastic case, where it causes collapse.  

The underlying reason is a qualitative difference in the evolution of the velocity dispersion profile. If central dissipation is sufficiently strong, the core does not become isothermal; instead, the positive velocity dispersion gradient persists at the center. As a consequence, conduction remains directed inward and counteracts central dissipation, which ultimately drives the collapse. In outer halo regions, where dissipation also dominates over outward conduction, we observe mass infall rather than outflow. In some cases, this leads to more enclosed mass at large radii than is achievable without dissipation and can generally increase the mass of a potentially forming central object.

This may be the most distinct signature of dissipation. For moderate dissipation, however, the observables generally evolve very similarly, only faster. Using rescaling arguments, we could explicitly show that the observed mass distribution of the recently discovered compact object in JVAS~B1938+666 can be explained by both an elastic and a weakly dissipative SIDM halo. In the late stages of their evolution, they generically develop steep slopes in the density profile, consistent with the observation. With dissipation, the required evolution time can be reduced by roughly a factor of two, or equivalently, a smaller cross section suffices.

Dissipation can therefore provide a natural mechanism to evade constraints on the self-interaction cross section. Realistically, both the cross section, \(\sigma_\mathrm{T}/m_\chi\), and \(r_\mathrm{diss}\) depend on velocity and are explicitly connected to a particle physics model. An emerging angular dependence may require a dissipative generalization of the hybrid approach between fSIDM and rSIDM \citep{Arido2025}. To generalize the conclusions of this work, initial conditions outside the rescaling family considered here should be explored, along with methods to continue simulations deeper into the collapse phase. Finally, cosmological simulations of dissipative SIDM will be essential to deriving observational constraints on the dissipation. Taken together, our results motivate a broader theoretical and observational program to constrain the dissipative properties of DM and their implications for particle physics models of the dark sector.

\begin{acknowledgements}
The authors thank all participants of the Darkium SIDM Journal Club for helpful discussions. We acknowledge support by the DFG Collaborative Research Institution Neutrinos and Dark Matter in Astro- and Particle Physics (SFB 1258) and the Excellence Cluster ORIGINS -- EXC-2094 -- 390783311.
MSF gratefully acknowledges the support of the Alexander von Humboldt Foundation through a Feodor Lynen Research Fellowship. 
We thank Kimberly Boddy for publishing the \textsc{GravothermalSIDM} code.
Other software:
NumPy \citep{numpy},
Matplotlib \citep{matplotlib}.
Colossus \citep{Diemer2018}.
\end{acknowledgements}

\noteadded{During the completion of this work, \citet{Zhang2026} appeared, discussing how elastic SIDM can account for the density profile of the perturber in the strong lens system JVAS~B1938+666 inferred by \citet{Vegetti2026}. Our results in \cref{sec:VegettiObservation} regarding the same system are in broad qualitative agreement for the elastic scenario.}

\bibliographystyle{aa}
\bibliography{references}

\begin{appendix}
\nolinenumbers

\section{Cooling and heat conduction rate}\label{appsec:coolingAndCondRate}
In this section, we compute the energy change of individual numerical particles per time step. After thermal averaging, this yields the cooling and heat conduction rates simulated by our \(N\)-body algorithm.

Using the velocity updates from \cref{eq:velocityUpdate} and the relative orientation of the involved vectors,
\begin{equation}
    \boldsymbol{v}_i - \boldsymbol{v}_j = \Delta \boldsymbol{v}_{ij} \parallel \Delta \boldsymbol{v}_\mathrm{drag} \perp \Delta \boldsymbol{v}_\mathrm{rand},
\end{equation}
the respective change in kinetic energy is given by
\begin{equation}\label{eq:DeltaEkin}
\begin{split}
    \frac{2\Delta E_{i/j}}{m} &= \boldsymbol{v}'^2_{i/j} - \boldsymbol{v}^2_{i/j} \\
    &= \Delta \boldsymbol{v}^2_\mathrm{drag} + \Delta \boldsymbol{v}^2_\mathrm{rand} \mp 2 \boldsymbol{v}_{i/j}\cdot \Delta \boldsymbol{v}_\mathrm{drag} \pm 2 \boldsymbol{v}_{i/j}\cdot \Delta \boldsymbol{v}_\mathrm{rand}\\
    &=\Delta v_\mathrm{drag} \frac{\Delta v_{ij}}{r_\mathrm{diss}} \mp 2\boldsymbol{v}_{i/j} \cdot\Delta\hat{\boldsymbol{v}}_{ij} \Delta v_\mathrm{drag} \pm 2 \Delta\boldsymbol{v}_{i/j} \cdot\Delta\hat{\boldsymbol{v}}_\mathrm{rand} \Delta v_\mathrm{rand}
\end{split}
\end{equation}
with
\begin{equation}
    \Delta v_\mathrm{drag} = \frac{1}{2}\Delta v_{ij}^2 \frac{\sigma_\mathrm{T}}{m_\chi} r_\mathrm{diss}\frac{\Delta t}{m}\int \mathrm{d}V\,\rho_i\rho_j.
\end{equation}
The upper sign always corresponds to the first index \(i\), the lower sign to the second index \(j\). In the second line of \cref{eq:DeltaEkin}, we inserted \(\Delta v_\mathrm{rand}^2\) from \cref{eq:RandKick}, which cancels \(\Delta v_\mathrm{drag}^2\). Taking the expectation value, the last term vanishes since it is odd under \(\Delta\hat{\boldsymbol{v}}_\mathrm{rand} \to - \Delta\hat{\boldsymbol{v}}_\mathrm{rand}\); the two have the same probability of being sampled from the uniform distribution of vectors in the plane perpendicular to \(\Delta \boldsymbol{v}_{ij}\). 

The local volumetric rate of energy change is found by thermally averaging, removing the volume integral, and dividing by the numerical time step, \(\Delta t\):
\begin{equation}\label{eq:RateOfEnergyChange}
    \frac{\mathrm{d}E_{i/j}}{\mathrm{d}V\,\mathrm{d}t} \simeq \dv{}{V}\frac{\langle\Delta E_{i/j}\rangle}{\Delta t} = \frac{1}{4}\frac{\sigma_\mathrm{T}}{m_\chi}\rho_i\rho_j\left[\langle\Delta v_{ij}^3\rangle \mp 2r_\mathrm{diss}\langle\boldsymbol{v}_{i/j}\cdot \Delta \boldsymbol{v}_{ij} \Delta v_{ij}\rangle\right]
.\end{equation}
In this step, we transitioned from discrete density kernels of numerical particles to local continuous density fields for components \(i\) and \(j\). This corresponds to the limit of many numerical particles of fixed kernel size within a small integration volume, so their overlapping kernels form two locally homogeneous but potentially distinct density fields. As in the main text, we assume a velocity-independent cross section (\(\sigma_\mathrm{T}\)) and dissipation parameter (\(r_\mathrm{diss}\)) here. Otherwise, the thermal averaging would need to include these quantities, which are highly model-dependent. 

If both \(\boldsymbol{v}_i\) and \(\boldsymbol{v}_j\) follow Maxwell-Boltzmann distributions of 1D velocity dispersions \(\nu_i\) and \(\nu_j\), respectively, the magnitude of the relative velocity is distributed as
\begin{equation}
    f(\Delta v_{ij}) = \sqrt{\frac{2}{\uppi}}\frac{\Delta v_{ij}^2}{\left(\nu_i^2 + \nu_j^2\right)^{3/2}} \exp\left(-\frac{\Delta v_{ij}^2}{2(\nu_i^2 + \nu_j^2)}\right).
\end{equation}
This is also a Maxwell-Boltzmann distribution with velocity dispersion \(\left(\nu_i^2 + \nu_j^2\right)^{1/2}\). The integral 
\begin{equation}
    \langle\Delta v_{ij}^3\rangle = \int\limits_0^\infty \mathrm{d}(\Delta v_{ij})\,\Delta v_{ij}^3 f(\Delta v_{ij}) = 8\sqrt{\frac{2}{\uppi}}\left(\nu_i^2 + \nu_j^2\right)^{3/2}
\end{equation}
can be evaluated analytically. 

For the other average in \cref{eq:RateOfEnergyChange}, we integrated separately over all velocity components, each of which follows a Gaussian,
\begin{equation}
\begin{split}
    &\langle\boldsymbol{v}_{i/j}\cdot \Delta \boldsymbol{v}_{ij} \Delta v_{ij}\rangle = \int \dd{\boldsymbol{v}_i} \dd{\boldsymbol{v}_j} \frac{\exp\left(-\frac{\boldsymbol{v}_i^2}{2\nu_i^2}\right)\exp\left(-\frac{\boldsymbol{v}_j^2}{2\nu_j^2}\right)}{(2\uppi)^3 \nu_i^3 \nu_j^3}\boldsymbol{v}_{i/j} \cdot \Delta \boldsymbol{v}_{ij} \Delta v_{ij}\\
    &= - \nu_{i/j}^2 \int \dd{\boldsymbol{v}_i} \dd{\boldsymbol{v}_j} \left(\nabla_{\boldsymbol{v}_{i/j}}\frac{\exp\left(-\frac{\boldsymbol{v}_i^2}{2\nu_i^2}\right)\exp\left(-\frac{\boldsymbol{v}_j^2}{2\nu_j^2}\right)}{(2\uppi)^3 \nu_i^3 \nu_j^3}\right) \cdot \Delta \boldsymbol{v}_{ij} \Delta v_{ij}, 
\end{split}
\end{equation}
and absorbed \(\boldsymbol{v}_{i/j}\) in a gradient, which produces the case-dependent \(\nu_{i/j}^2\) in front of the integral. Integrating by parts and employing that the Gaussians vanish for \(\boldsymbol{v}_{i/j}\to\infty\), we find the relation 
\begin{equation}
    \langle\boldsymbol{v}_{i/j}\cdot \Delta \boldsymbol{v}_{ij} \Delta v_{ij}\rangle = \nu_{i/j}^2\left\langle\nabla_{v_{i/j}}\left(\Delta \boldsymbol{v}_{ij} \Delta v_{ij}\right)\right\rangle = \pm 4\nu_{i/j}^2\left\langle\Delta v_{ij}\right\rangle 
\end{equation}
by direct differentiation.

After evaluating
\begin{equation}
    \langle\Delta v_{ij}\rangle = \int\limits_0^\infty \mathrm{d}(\Delta v_{ij})\,\Delta v_{ij} f(\Delta v_{ij}) = 2\sqrt{\frac{2}{\uppi}}\left(\nu_i^2 + \nu_j^2\right)^{1/2},
\end{equation}
\cref{eq:RateOfEnergyChange} reduces to
\begin{equation}\label{eq:CoolingAndConduction}
\begin{split}
    \frac{\mathrm{d}E_{i/j}}{\mathrm{d}V\,\mathrm{d}t} &= \sqrt{\frac{8}{\uppi}}\frac{\sigma_\mathrm{T}}{m_\chi}\rho_i \rho_j \left[(\nu_i^2 + \nu_j^2)^{3/2} - 2r_\mathrm{diss}\nu_{1/2}^2 (\nu_i^2 + \nu_j^2)^{1/2}\right]\\
    &= -\sqrt{\frac{8}{\uppi}}\frac{\sigma_\mathrm{T}}{m_\chi} (r_\mathrm{diss} - 1) \rho_i \rho_j (\nu_i^2 + \nu_j^2)^{3/2}\\  
    &\mp\sqrt{\frac{8}{\uppi}}\frac{\sigma_\mathrm{T}}{m_\chi} r_\mathrm{diss}\rho_i\rho_j(\nu_i^2 + \nu_j^2)^{1/2}(\nu_i^2 - \nu_j^2).
\end{split}
\end{equation}
The result naturally separates into a dissipation term, which vanishes for \(r_\mathrm{diss} = 1\), and a conduction term, which vanishes for \(\nu_i = \nu_j\). Note that this is the volumetric rate of energy change in scattering processes between two components, \(i\) and \(j\), that do not necessarily have to differ in density and velocity dispersion. Self-scattering within a component gives rise to the local cooling rate
\begin{equation}
    C(\rho,\nu) = \frac{8}{\sqrt{\uppi}}\frac{\sigma_\mathrm{T}}{m_\chi}(r_\mathrm{diss} - 1)\rho^2\nu^3
\end{equation}
by inserting \(\rho\equiv \rho_i = \rho_j\) and \(\nu\equiv \nu_i = \nu_j\) into \cref{eq:CoolingAndConduction}. For multiple components, all pairwise combinations must be taken into account.

\section{Validation of the fSIDM algorithm}\label{sec:Validation}
We validated the cooling and heat conduction rate in our implementation of the dissipative fSIDM algorithm against \cref{eq:CoolingAndConduction} by simulating a periodic box containing one or two particle components of homogeneous density and velocity dispersion in the absence of gravity. Since the density remains constant, the evolution is governed by differential equations for the velocity dispersion that are analytically or at least numerically tractable and should be reproduced by the \(N\)-body simulation.

\subsection{Test with dissipation only}
\begin{figure*}[ht!]
    \centering
    \includegraphics[width=\hsize]{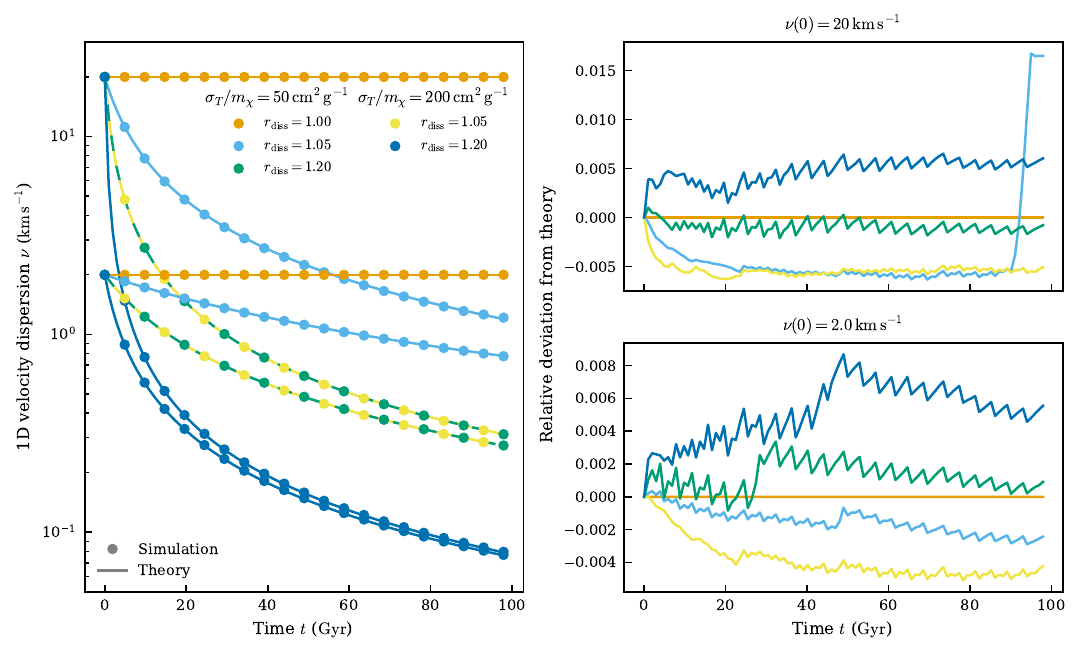}
    \caption{\textit{Left:} Comparison of the simulated velocity dispersion (dots, shown every \(5\,\mathrm{Gyr}\)) in a periodic box of homogeneous density with the theoretical expectation (lines) for different cooling rates set by the modified transfer cross section \(\sigma_\mathrm{T}/m_\chi\) (see Eq. \ref{eq:TransferCrossSection}) and dissipation parameter \(r_\mathrm{diss}\). 
    \textit{Right:} Relative deviation of the simulation from the theoretical prediction for the two initial velocity dispersions. With a single exception at late times, the agreement is at the sub-percent level.}
    \label{fig:dissipationTest}
\end{figure*}
To examine the cooling rate, we initialized \(N = 10^4\) particles with a total mass of \(M = 10^{10} \,\mathrm{M}_\odot\) in a periodic cube of side length \(L = 10\,\mathrm{kpc}\). We considered two initial 1D velocity dispersions, \(\nu(t=0) = 20\,\mathrm{km}\,\mathrm{s}^{-1}\) and \(\nu(t=0) = 2.0\,\mathrm{km}\,\mathrm{s}^{-1}\), and varied the velocity-independent parameters \(\sigma_\mathrm{T}/m_\chi\) and \(r_\mathrm{diss}\) \citep[similar setup as in][Appendix B.3]{Shen2024}. Initial positions are drawn from a uniform distribution, while velocity components are sampled from a zero-mean Gaussian with standard deviation \(\nu(t=0)\), which corresponds to a Maxwell-Boltzmann distribution. 

Assuming that the velocities remain Maxwell-Boltzmann distributed, the evolution of the velocity dispersion in the simulation should be governed by
\begin{equation}\label{eq:cooling}
    \frac{3}{2}\rho\dv{\nu^2}{t} = - C(\rho,\nu) = -\frac{8}{\sqrt{\uppi}} \frac{\sigma_\mathrm{T}}{m_\chi}(r_\mathrm{diss} - 1)\rho^2\nu^3
,\end{equation}
where we used the local cooling rate \(C(\rho,\nu)\) derived in Appendix~\ref{appsec:coolingAndCondRate}. This has the analytic solution
\begin{equation}\label{eq:DissipationTestSolution}
    \nu(t) = \nu(t=0)\left(1+\frac{t}{2t_\mathrm{diss}}\right)^{-1}
\end{equation}
with the dissipation timescale
\begin{equation}\label{eq:DissipationTimescales}
    t_\mathrm{diss} = \frac{3\rho\nu(t=0)^2}{2C(\rho,\nu(t=0))} = \frac{3\sqrt{\uppi}}{16\frac{\sigma_\mathrm{T}}{m_\chi}(r_\mathrm{diss} - 1)\rho\nu(t=0)}.
\end{equation}

Due to finite particle sampling, the average velocity, \(\langle\boldsymbol{v}\rangle\), in the simulation's initial conditions does not vanish exactly. Since it is conserved in the simulation, it becomes dominant once the velocity dispersion has decreased to \(\nu(t) \sim |\langle\boldsymbol{v}\rangle|\). However, the analytic solution assumes \(\langle\boldsymbol{v}\rangle = 0\) and thus overestimates the relative velocity in interactions and therefore dissipation. To mitigate this, we subtracted the mean velocity from all particles to enforce \(\langle\boldsymbol{v}\rangle = 0\) while preserving \(\nu(0)\).

In \cref{fig:dissipationTest} we compare the simulated time evolution of the velocity dispersion with the analytic expectation. The left panel shows that the velocity dispersion decays toward zero at a rate depending on \(\sigma_\mathrm{T}/m_\chi(r_\mathrm{diss}-1)\), as predicted by \cref{eq:DissipationTestSolution,eq:DissipationTimescales}. At late times \(t\gg t_\mathrm{diss}\), the evolution becomes independent of the initial velocity dispersion. Simulations with identical \(\sigma_\mathrm{T}/m_\chi(r_\mathrm{diss} - 1)\) (green and yellow) yield indistinguishable results, confirming the expected degeneracy. The right panels show the relative deviation of the simulation from the analytic solution and demonstrate sub-percent-level agreement. Residual high-frequency fluctuations arise primarily from snapshots written when not all particles have completed their current time step. In summary, the test confirms that the implementation of dissipative interactions reproduces the expected cooling rate in a controlled setting without gravity.

\subsection{Test with additional heat conduction}
\begin{figure*}[ht!]
    \centering
    \includegraphics[width=\hsize]{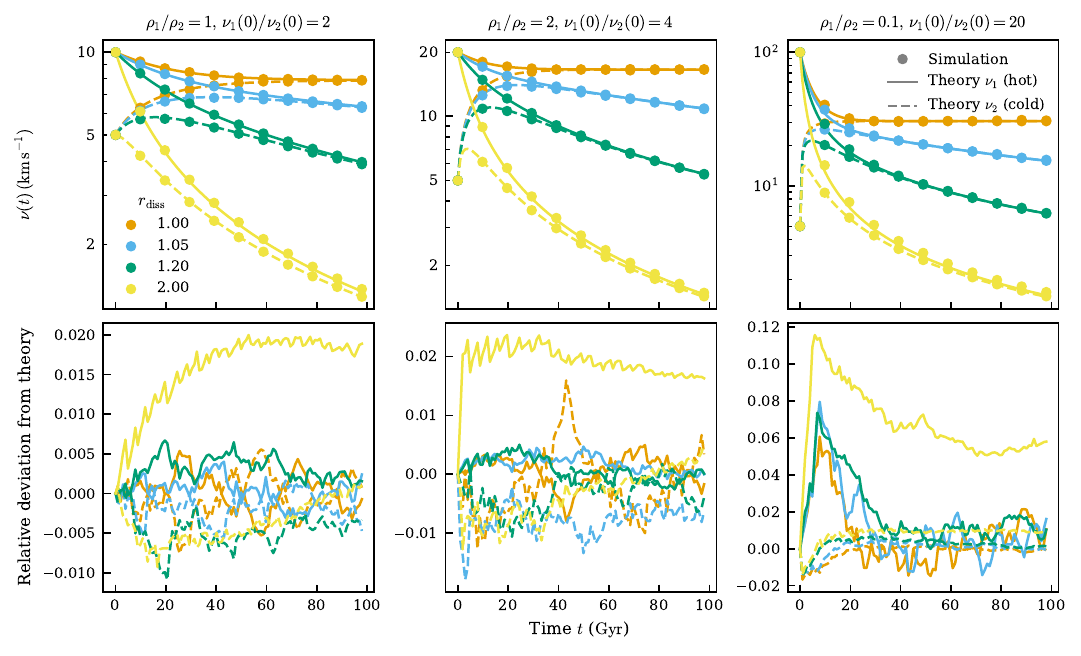}
    \caption{\textit{Top row:} Comparison of the simulated velocity dispersions (dots, shown every \(10\,\mathrm{Gyr}\)) with the numerical solution of \cref{eq:TwoComponentODE} (lines) for two components of different initial velocity dispersions and densities. The cooler one (dashed) always starts at \(\nu(0) = 5\,\mathrm{km}\,\mathrm{s}^{-1}\), and the cross section is set to \(\sigma_\mathrm{T}/m_\chi = 2\,\mathrm{cm}^2\,\mathrm{g}^{-1}\) in all cases.
    \textit{Bottom row:} Relative deviation of the simulation results from the theoretical prediction. Solid lines correspond to the hotter component and dashed lines to the cooler component.}
    \label{fig:twoComponentTest}
\end{figure*}
To test a system subject to both dissipation and heat conduction, we increased the number of particles to \(N = 2\times 10^4\) and divided them into two components with different initial velocity dispersions, \(\nu_1(t=0) > \nu_2(t=0)\), and relative abundance \(f = \rho_1/\rho_2\). The total mass \(M = 10^{10} \,\mathrm{M}_\odot\) and box size \(L = 10\,\mathrm{kpc}\) are unchanged, and gravity remains disabled. 

The time evolution of the two velocity dispersions is governed by coupled differential equations for each component \(i=1,2\) with \(i\neq j\)
\begin{equation}\label{eq:TwoComponentODE}
\begin{split}
    \frac{3}{2}\rho_i\frac{\mathrm{d}\nu_i^2}{\mathrm{d}t} = &-\frac{8}{\sqrt{\uppi}} \frac{\sigma_\mathrm{T}}{m_\chi}(r_\mathrm{diss} - 1)\rho_i^2\nu_i^3\\ &- \frac{8}{\sqrt{\uppi}} \frac{\sigma_\mathrm{T}}{m_\chi}(r_\mathrm{diss} - 1)\rho_i\rho_j\left(\frac{\nu_i^2 + \nu_j^2}{2}\right)^{3/2}\\ &- \sqrt{\frac{8}{\uppi}}\frac{\sigma_\mathrm{T}}{m_\chi}r_\mathrm{diss}\rho_i\rho_j\left(\nu_i^2 + \nu_j^2\right)^{1/2}\left(\nu_i^2 - \nu_j^2\right),
\end{split}
\end{equation}
which are derived in Appendix~\ref{appsec:coolingAndCondRate}. The first two terms are due to cooling by scatterings within the component and between the components. The last term is due to conduction and transports energy from the hotter to the cooler component, as it is antisymmetric under \(\nu_i\leftrightarrow\nu_j\). It closely resembles the conduction term found by \citet{Dvorkin2014} and used by \citet{Fischer2025},\footnote{Since they consider DM-baryon scatterings, they use the usual momentum transfer cross section for distinguishable particles instead of the modified one introduced in \cref{eq:TransferCrossSection}.} but is enhanced by a factor \(r_\mathrm{diss}\). 

We fixed the cross section at \(\sigma_\mathrm{T}/m_\chi = 2\,\mathrm{cm}^2\,\mathrm{g}^{-1}\), as a rescaling of \(\sigma_\mathrm{T}/m_\chi\) is equivalent to a rescaling of time \(t\to t/(\sigma_\mathrm{T}/m_\chi)\). The system further exhibits a scaling symmetry under \(\nu_{i/j}\to\lambda\nu_{i/j}\) and \(t \to t/\lambda\). This implies that the dynamics depend only on the ratio of the initial velocity dispersions \(\nu_1(t=0)/\nu_2(t=0)\), not the absolute scale. We therefore fixed \(\nu_2(t=0) = 5\,\mathrm{km}\,\mathrm{s}^{-1}\). Similarly, the equations are invariant under \(\rho_{i/j}\to\lambda\rho_{i/j}\) and \(t \to t/\lambda\), implying that the evolution depends only on the relative abundance \(f = \rho_1/\rho_2\). Larger cross sections, velocities and densities therefore accelerate the evolution, consistent with the scaling of the scattering rate \(\propto \rho\nu\sigma_\mathrm{T}/m_\chi\).

\Cref{fig:twoComponentTest} shows the simulated time evolution of the velocity dispersions compared to the numerical solution of \cref{eq:TwoComponentODE} for several representative initial configurations (columns) and dissipation parameters \(r_\mathrm{diss}\) (colors). There are two competing processes: heat conduction drives the two components toward a common velocity dispersion, and cooling (for \(r_\mathrm{diss}>1\)) makes the velocity dispersion decay. Increasing \(r_\mathrm{diss}\) accelerates this decay. The relative difference between simulation and theory generally increases with dissipation, as a similar absolute difference appears larger relative to the smaller velocity dispersions. Initial configurations with a larger contrast between the two components also show larger discrepancies. 

Regardless of whether the hotter component is denser or less dense than the cooler one, heat transfer appears to proceed slightly more slowly than predicted by \cref{eq:TwoComponentODE}. We found that this discrepancy can be reduced to some extent by adopting smaller time steps to better resolve the rapid initial conduction phase. Even with very small time steps, there may be a residual deviation if individual interactions with large dissipation or heat transfer drive the velocity distribution away from Maxwell-Boltzmann. Nevertheless, the overall agreement for the simulations shown here is very good.

\section{Convergence tests}\label{appsec:ConvergenceTest}
A fundamental problem for our dissipative \(N\)-body simulation is the realization noise arising from the discrete representation of the analytic NFW halo. In particular, the initial energy varies on the percent-level around the analytic initial energy \(\hat{E}(0) = -392.85\times 10^{10}\, \mathrm{M}_\odot\,\mathrm{km}^2\,\mathrm{s}^{-2}\). 

\begin{figure}
    \centering
    \includegraphics[width = \hsize]{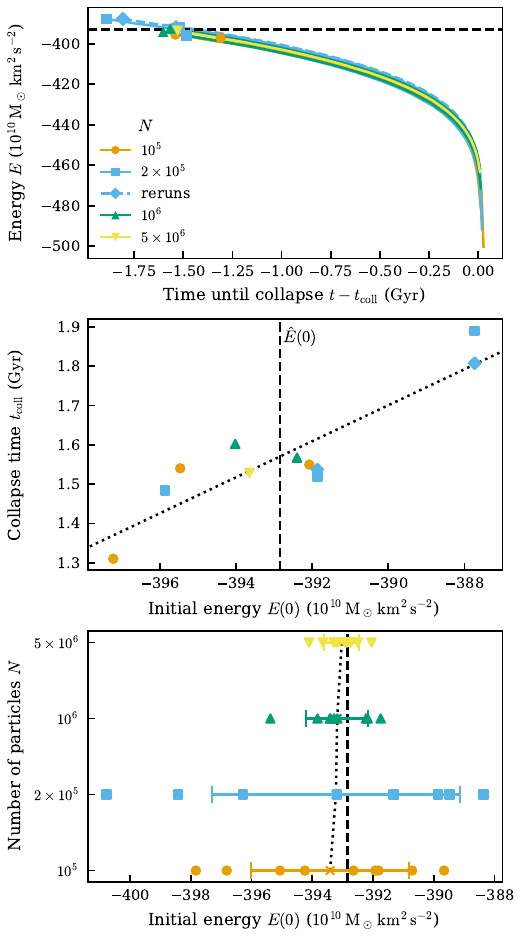}
    \caption{\textit{Top:} Total energy as a function of time until collapse for different discrete realizations of the same isolated halo. \textit{Middle:} Correlation between the collapse time and initial energy with a linear fit (dotted). \textit{Bottom:} Scatter in the initial energy of nine sampled initial conditions for each number of particles, \(N\). The error bars mark one standard deviation, and the dotted line connects the means (crosses). In all panels, the dashed lines mark the desired value of the initial energy, \(\hat{E}(0) = -392.85\times 10^{10}\, \mathrm{M}_\odot\,\mathrm{km}^2\,\mathrm{s}^{-2}\), around which the discrete initial conditions are scattered due to realization noise.}
    \label{fig:Einit_tcoll}
\end{figure}

For dissipative simulations with \(\sigma_\mathrm{T}/m_\chi = 50\,\mathrm{cm}^2\,\mathrm{g}^{-1}\) and \(r_\mathrm{diss} = 1.05\), this variation is significantly amplified in the collapse time, as shown in \cref{fig:Einit_tcoll}. Configurations of lower initial energy collapse earlier, while those of higher initial energy take longer to collapse. The top panel shows the energy as a function of time remaining until collapse. All simulations follow a very similar trajectory, which is initially linear to a good approximation. For the simulation parameters adopted, the collapse consistently occurs near a fixed total energy. Reruns with identical initial conditions yield very similar results, confirming that the variation is dominated by the initial conditions rather than the inherent stochasticity of the algorithm.

The middle panel can be considered as focusing on the region with markers in the top panel with inverted axes. Due to the initially linear evolution of the total energy, the correlation between collapse time and initial energy is well described by a two-parameter fit:
\begin{equation}
    t_\mathrm{coll} = \left[(0.046\pm0.008) \frac{E(0) - \hat{E}(0)}{10^{10}\,\mathrm{M}_\odot\,\mathrm{km}^2\,\mathrm{s}^{-2}} + (1.57\pm0.02)\right]\,\mathrm{Gyr}.
\end{equation}
Thus, a 1\,\%-variation in initial energy translates into a much larger 18\,\%-variation in the collapse time. This relation is independent of \(N\); however, the typical deviation from the desired \(\hat{E}(0)\) decreases with \(N\), as shown in the bottom panel. The mean slightly underestimates \(\hat{E}(0)\) systematically.

In the following, we use the realization closest to \(\hat{E}(0)\) among nine generated initial conditions for each \(N\). Their relative deviations from \(\hat{E}(0)\) range from \(8.9\times10^{-4}\) for \(N = 2\times10^5\) down to \(2.8\times10^{-5}\) for \(N = 5\times 10^6\). 

\begin{figure*}
    \centering
    \includegraphics[width = \hsize]{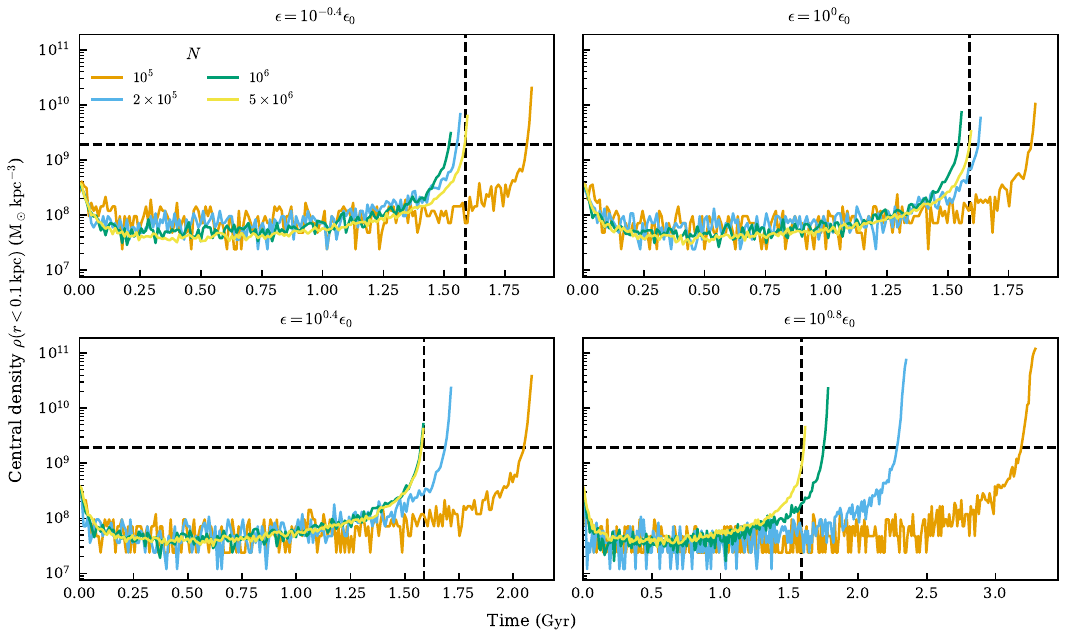}
    \caption{Time evolution of the central density varying the number of simulation particles (\(N\)) and the softening length (\(\epsilon\)) around the reference value, \(\epsilon_0\) (see Eq. \ref{eq:SofteningLength}). For reference, dashed lines mark the collapse time (\(t_\mathrm{coll}\)) for \(\epsilon=\epsilon_0\) and \(N = 5\times 10^6\) and the corresponding density, \(\rho = 5\rho(t=0)\).}
    \label{fig:softeningTest}
\end{figure*}

With these initial conditions, we investigated the impact of the gravitational softening length, \(\epsilon\), on the simulation results, following \citet{Mace2024}. It regularizes the gravitational potential 
\begin{equation}
    \Phi(\boldsymbol{r}) = -\mathrm{G}\sum\limits_{i=1}^N\frac{m}{\left[(\boldsymbol{r} - \boldsymbol{r}_i)^2 + \epsilon^2\right]^{1/2}}
\end{equation}
generated by the simulation particles, preventing infinities, large-angle gravitational scatterings, the formation of binaries, and prohibitively small time steps \citep{Springel2016}.

We varied the softening length around the reference value
\begin{equation}\label{eq:SofteningLength}
    \epsilon_0 = \frac{2.3r_\mathrm{s}}{\sqrt{N(r<r_\mathrm{s})}}\simeq\begin{cases}
        0.09\,\mathrm{kpc}, & N = 10^5\\
        0.06\,\mathrm{kpc}, & N = 2\times10^5\\
        0.03\,\mathrm{kpc}, & N = 10^6\\
        0.01\,\mathrm{kpc}, & N = 5\times10^6\\
    \end{cases}
\end{equation}
with NFW scale radius, \(r_\mathrm{s}\), and the number of particles, \(N(r <r_\mathrm{s})\), enclosed within. It is inspired by the criterion from \citet{Power2003}, but is 6--7 times smaller \citep[motivated by][]{Fischer2025a}, and was also used for \cref{fig:Einit_tcoll} and throughout the main text. 

\Cref{fig:softeningTest} shows the time evolution of the central density varying the softening length around \(\epsilon_0\) for different numbers of simulation particles, \(N\). The simulation parameters \(\sigma_\mathrm{T}/m_\chi = 50\,\mathrm{cm}^2\,\mathrm{g}^{-1}\) and \(r_\mathrm{diss} = 1.05\) are unchanged compared to \cref{fig:Einit_tcoll} but we now use the most accurate initial conditions we found. For small \(N\), a large softening length systematically delays core collapse. It flattens not only the potential of every single particle, but also that of the full halo \citep{Barnes2012}. The central mass subject to dissipation may therefore concentrate less, which in turn reduces dissipation. 

Our largest simulation with \(N = 5\times10^6\) is almost insensitive to the softening length. For this reason, and due to the reduced realization noise, which is also evident in the density evolution, we ran all simulations considered in the main text with this particle number.

\section{The gravothermal fluid model with dissipation}\label{appsec:FluidModel}
The gravothermal fluid model \citep{Gnedin2001,Balberg2002,Koda2011} considers isolated SIDM halos as quasi-static, spherically symmetric systems of a conductive, monatomic ideal gas (adiabatic index \(\gamma = 5/3\)) with pressure \(P = \rho k_B T/m_\chi = \rho\nu^2\). It was first introduced for globular clusters \citep{LyndenBell1980}; \citet{Bettwieser1986} even formulated it with an anisotropic velocity dispersion \(\nu\). Nevertheless, the standard set of equations for SIDM assumes isotropy:
\begin{align}
    \pdv{M}{r} &= 4\uppi r^2\rho\label{eq:Continuity}\\
    \pdv{(\rho\nu^2)}{r} &= -\frac{\mathrm{G}M\rho}{r^2}\label{eq:HydrostaticEquilibrium}\\
    \rho\nu^2\left(\dv{}{t}\right)_M \log \frac{\nu^3}{\rho} &= -\frac{1}{4\uppi r^2}\pdv{L}{r} - C(\rho,\nu)\label{eq:EnergyChange}\\
    \frac{L}{4\uppi r^2} &= -\frac{m_\chi}{k_B} \kappa \pdv{\nu^2}{r}\label{eq:HeatConduction},
\end{align}
Together, they describe the conservation of mass (Eq. \ref{eq:Continuity}), the hydrostatic equilibrium between pressure and gravity (Eq. \ref{eq:HydrostaticEquilibrium}), the change of energy density within each mass element (Eq. \ref{eq:EnergyChange}), and the luminosity, \(L\), due to heat conduction down the gradient of the velocity dispersion (Eq. \ref{eq:HeatConduction}). The heat conductivity is conventionally interpolated as a harmonic mean, \(\kappa^{-1} = \kappa_\mathrm{SMFP}^{-1} + \kappa_\mathrm{LMFP}^{-1}\), between the SMFP limit
\begin{equation}
    \kappa_\mathrm{SMFP} = \frac{75\sqrt{\uppi}\nu}{64\sigma_\mathrm{tot}}
\end{equation}
and the LMFP limit
\begin{equation}
    \kappa_\mathrm{LMFP} = \frac{3\beta}{2\uppi^{3/2} \mathrm{G}}\frac{\sigma_\mathrm{tot}}{m_\chi}\rho\nu^3.
\end{equation}
Motivated by Sect.~\ref{sec:fSIDMrSIDMmatching}, we identify the total cross section with the transfer cross section of the fSIDM algorithm via \(\sigma_\mathrm{tot} = 3\sigma_\mathrm{T}\). 

\begin{figure}
    \centering
    \includegraphics[width = \hsize]{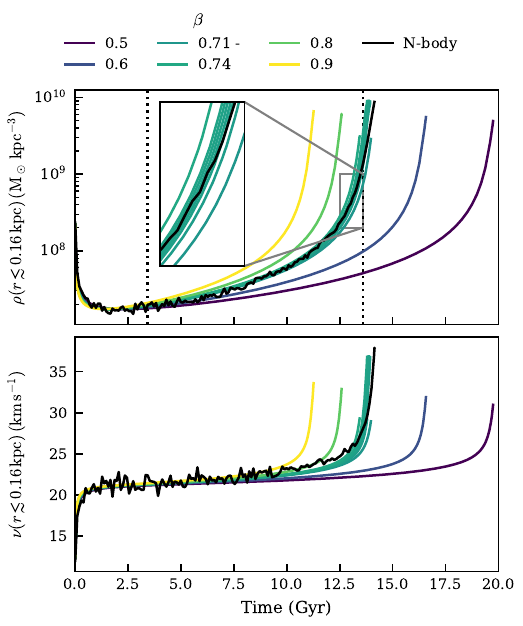}
    \caption{Time evolution of the central density and velocity dispersion obtained from the elastic fluid model for different values of the calibration parameter, \(\beta\). The collapse time increases considerably with decreasing \(\beta\), i.e., with decreasing LMFP heat conduction. Between the two vertical dotted lines, \(\beta = 0.726\) minimizes the deviation from the \(N\)-body central density (black curve).}
    \label{fig:calibration}
\end{figure}

The LMFP limit contains an \(\order{1}\) parameter \(\beta\) to be calibrated. For this purpose, we employed the non-dissipative (\(r_\mathrm{diss} = 1.0\)) \(N\)-body simulation with \(\sigma_\mathrm{T}/m_\chi = 50\,\mathrm{cm}^2\,\mathrm{g}^{-1}\) and considered the central density (\cref{fig:calibration}, top panel). The smaller \(\beta\), i.e., the weaker the LMFP conduction, the later the core collapse. We find the best agreement in terms of time-averaged squared relative deviation
\begin{equation}
    \mathcal{L} = \left\langle\left(\frac{\rho_\mathrm{Fluid} - \rho_\mathrm{Nbody}}{\rho_\mathrm{Fluid}}\right)^2\right\rangle_{t \in [t_\mathrm{core} + 0.1t_\mathrm{coll}, t_\mathrm{coll}]}
\end{equation}
for \(\beta = 0.726\). The core formation phase is excluded from the fit, as the fluid model does not reproduce it well for any value of \(\beta\).

\Cref{eq:EnergyChange} includes the cooling rate, \(C(\rho,\nu)\), \citep[cf.][]{Essig2019} to account for dissipation, as well as a Lagrangian derivative,
\begin{equation}
    \left(\dv{}{t}\right)_M = \pdv{}{t} + \boldsymbol{v}\cdot\nabla
,\end{equation}
which follows the mass elements moving at a velocity \(\boldsymbol{v}\) rather than taking the time derivative at a fixed position. In practice, the \textsc{GravothermalSIDM} code \citep{Nishikawa2020,Outmezguine2023,GadNasr2024} discretizes the halo into shells of constant mass, whose energies are updated according to \cref{eq:EnergyChange,eq:HeatConduction}. Hydrostatic equilibrium is subsequently restored by shifting the mass shells according to \cref{eq:Continuity,eq:HydrostaticEquilibrium}.

\section{Matching between fSIDM and rSIDM}\label{appsec:Matching}
\begin{figure}
    \centering
    \includegraphics[width = \hsize]{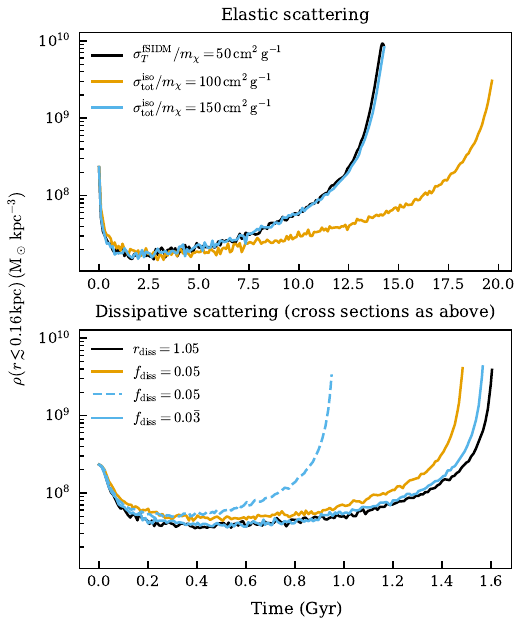}
    \caption{Central density evolution of representative elastic (\textit{top}) and dissipative (\textit{bottom}) fSIDM simulations (black curves) to be matched by rSIDM simulations (colored curves). The legend in the top panel applies to both panels; the legend in the bottom panel applies only there. In the elastic case, the matching \(\sigma_\mathrm{tot}^\mathrm{iso} \leftrightarrow 3\sigma_\mathrm{T}^\mathrm{fSIDM}\) is clearly superior. With dissipation, the dissipation parameter for rSIDM should additionally satisfy \(f_\mathrm{diss}\leftrightarrow 2(r_\mathrm{diss}-1)/3\).}
    \label{fig:Matching}
\end{figure}

We ran two elastic and three dissipative velocity-independent isotropic rSIDM simulations with the same initial conditions as used for fSIDM (see~Sect.~\ref{sec:Setup}) to verify the matching condition given in \cref{eq:fSIDMrSIDMmatching}. It follows from requiring equal viscosity cross sections \citep{Yang2022} in the strongly forward-dominated and isotropic cases, respectively, together with equal cooling rates. \Cref{fig:Matching} clearly confirms the matching condition, demonstrating that the angular dependence of the cross section has only a limited influence on the evolution of the isolated halo. A more detailed discussion can be found in Sect.~\ref{sec:fSIDMrSIDMmatching}.  
\FloatBarrier

\section{Computing projected quantities}\label{appsec:ProjectedQuantities}
The projected surface density, \(\Sigma(r_\mathrm{2D})\), used in Sects.~\ref{sec:ProjectedQuantities} and \ref{sec:VegettiObservation} and the usual 3D density, \(\rho(r)\), are related by the Abel transform,
\begin{equation}
    \Sigma(r_\mathrm{2D}) \equiv \int\limits_{-\infty}^\infty \mathrm{d}z\, \rho(r) = 2\int\limits_{r_\mathrm{2D}}^\infty \mathrm{d}r\, \frac{r\rho(r)}{\sqrt{r^2 - r_\mathrm{2D}^2}}.
\end{equation}
For the numerical conversion of fluid model results, it is advantageous to retain the \(z\)-integral form near the singularity at \(r = r_\mathrm{2D}\). 

The projected enclosed mass directly follows as
\begin{equation}
    M(r<r_\mathrm{2D}) = 2\uppi \int\limits_0^{r_\mathrm{2D}} \mathrm{d}r\,r\Sigma(r).
\end{equation}
However, the fluid code generates density profiles only down to an inner radius \(r_0\), so the surface density is unavailable for \(r_\mathrm{2D} < r_0\). Therefore, we decomposed the projected enclosed mass into three contributions: the mass within a sphere of radius \(r_0\), the mass within the remaining cylinder of radius \(r_0\), and the mass within the annular cylinder between \(r_0\) and \(r_\mathrm{2D}\), where \(\Sigma(r_\mathrm{2D})\) is available. This yields
\begin{equation}
\begin{split}
    M(r<r_\mathrm{2D}) = M(r<r_0) &+ 4\uppi \int\limits_{r_0}^\infty \mathrm{d}r\,r\rho(r)\left(r-\sqrt{r^2 - r_0^2}\right)\\ 
    &+ 2\uppi\int\limits_{r_0}^{r_\mathrm{2D}}\mathrm{d}r\,r\Sigma(r).
\end{split}
\end{equation}
The logarithmic slope of the surface density profile is defined as 
\begin{equation}
    \gamma_\mathrm{2D}(r_\mathrm{2D}) = \dv{\log \Sigma}{\log r_\mathrm{2D}}.
\end{equation}
It is straightforward to compute numerically.

\section{Rescaling \(N\)-body results}\label{appsec:Rescaling}
Purely gravitational \(N\)-body simulations without self-interactions generate discrete realizations that sample the coarse-grained distribution function governed by the Poisson-Vlasov system
\begin{align}
    \nabla^2\Phi &= 4\uppi \mathrm{G} \int \mathrm{d}^3 v\, f(\boldsymbol{r},\boldsymbol{v},t),\\
    \dv{}{t} f(\boldsymbol{r},\boldsymbol{v},t) &= \pdv{f}{t} + \boldsymbol{v}\cdot\nabla_{\boldsymbol{r}} f - \nabla_{\boldsymbol{r}}\Phi\cdot\nabla_{\boldsymbol{v}} f = 0
\end{align}
with the gravitational potential \(\Phi\) and the distribution function \(f\) depending on position \(\boldsymbol{r}\), velocity \(\boldsymbol{v}\) and time \(t\). These equations are invariant under the two-parameter rescaling transformations 
\begin{equation}
\begin{split}
    &\boldsymbol{r}\to\lambda\boldsymbol{r},\quad \boldsymbol{v}\to\sqrt{\frac{\mu}{\lambda}}\boldsymbol{v},\quad t\to\sqrt{\frac{\lambda^3}{\mu}}t,\\&\Phi\to\frac{\mu}{\lambda}\Phi,\quad f\to\frac{1}{\sqrt{\mu\lambda^3}}f
\end{split}
,\end{equation}
where \(\mu\) rescales the macroscopic mass or, more fundamentally, the number of microscopic particles with constant mass \(m_\chi\). 

We can use the two parameters, \(\lambda\) and \(\mu\), to rescale the results of an \(N\)-body simulation, which are described by these invariant evolution equations. However, we must reinterpret the results as arising from different NFW initial conditions with \(r_\mathrm{s} \to \lambda r_\mathrm{s}\) and \(\rho_0\to\mu\rho_0/\lambda^3\) in order for the density profile
\begin{equation}
    \rho(r) = \frac{\rho_0}{r/r_\mathrm{s}\left(1+r/r_\mathrm{s}\right)^2}
\end{equation}
to remain invariant.

In the presence of self-interactions, the vanishing right-hand side of the Vlasov equation is formally replaced by the two-body collision term
\begin{equation}
\begin{split}
    \mathcal{C} = \int\mathrm{d}^3 v_2\int\mathrm{d}\Omega\, |\boldsymbol{v} - \boldsymbol{v}_2| \dv{\sigma}{\Omega}&\left(f(\boldsymbol{r},\boldsymbol{v}',t) f(\boldsymbol{r},\boldsymbol{v}'_2,t)\right.\\
    &\left.- f(\boldsymbol{r},\boldsymbol{v},t) f(\boldsymbol{r},\boldsymbol{v}_2,t)\right),
\end{split}
\end{equation}
which scatters velocities from \(\boldsymbol{v}\) to \(\boldsymbol{v}'\). Invariance of this term requires
\begin{equation}
    \dv{\sigma}{\Omega} \to \frac{\lambda^2}{\mu}\dv{\sigma}{\Omega}
\end{equation}
and similarly for all angular integrals such as \(\sigma_\mathrm{tot}\) or \(\sigma_\mathrm{T}\). Their dependence on the relative velocity must not introduce a fixed velocity scale arising from microphysics. This scaling symmetry is implicit in the dimensionless formulation of the gravothermal fluid equations \citep{LyndenBell1968,Balberg2002}, but to our knowledge has not been stated explicitly at the level of the Poisson-Vlasov system.

The dissipative generalization of the collision term reads
\begin{equation}
\begin{split}
    \mathcal{C} = \int\mathrm{d}^3 v_2\int\mathrm{d}\Omega\int \mathrm{d}k^0\, |\boldsymbol{v} - \boldsymbol{v}_2| \frac{\mathrm{d}\sigma}{\mathrm{d}\Omega\,\mathrm{d}k^0}&\left(f(\boldsymbol{r},\boldsymbol{v}',t) f(\boldsymbol{r},\boldsymbol{v}'_2,t)\right.\\ 
    &\left.- f(\boldsymbol{r},\boldsymbol{v},t) f(\boldsymbol{r},\boldsymbol{v}_2,t)\right)
\end{split}
,\end{equation}
with \(k^0 = m_\chi(\boldsymbol{v}^2 + \boldsymbol{v}_2^2 - \boldsymbol{v}'^2 - \boldsymbol{v}'^2_2)\). This implies \(k^0 \to \mu k^0/\lambda\) and 
\begin{equation}\label{eq:diffCrossSecRescaling}
    \frac{\mathrm{d}\sigma}{\mathrm{d}\Omega\,\mathrm{d}k^0} \to \frac{\lambda^3}{\mu^2} \frac{\mathrm{d}\sigma}{\mathrm{d}\Omega\,\mathrm{d}k^0}.
\end{equation}
Analogously to the velocity dependence in the elastic case, this differential cross section must be free of fixed scales to satisfy \cref{eq:diffCrossSecRescaling} for arbitrary \(k^0\). In particular, our simulations with velocity-independent transfer cross section and energy radiation proportional to the kinetic energy satisfy this condition. While \(\sigma_\mathrm{T}/m_\chi \to \lambda^2/\mu\, \sigma_\mathrm{T}/m_\chi\), the dimensionless dissipation parameter \(r_\mathrm{diss}\) remains invariant. 

In summary, rescaling simulation results by \(\lambda\) and \(\mu\) corresponds to evolving a rescaled NFW halo with a different cross section but the same dissipation parameter \(r_\mathrm{diss}\). We explicitly verified that our \(N\)-body implementation exhibits the rescaling symmetry by evolving an NFW halo of \(\rho_0 = 5.40\times10^7\,\mathrm{M}_\odot\,\mathrm{kpc}^{-3}\) and \(r_\mathrm{s} = 0.1944\,\mathrm{kpc}\) with \(\sigma_\mathrm{T}/m_\chi = 121.5\,\mathrm{cm}^2\,\mathrm{g}^{-1}\) and \(r_\mathrm{diss} = 1.01\). These parameters are related to our fiducial simulation of \(r_\mathrm{diss}=1.01\) by \(\lambda = 0.054\) and \(\mu = 0.0012\). Under this rescaling, all generated results are in perfect agreement. 

In Sect.~\ref{sec:VegettiObservation}, we mapped our simulations to agree with two observed data points \(\left(r^\mathrm{obs}_\mathrm{2D}, M(r<r^\mathrm{obs}_\mathrm{2D})\right)\) of a compact object for \(r_\mathrm{2D} = 20\,\mathrm{pc}\) and \(90\,\mathrm{pc}\). To this end, we rescaled distances as
\begin{equation}
    \lambda = \frac{r^\mathrm{obs}_\mathrm{2D}}{r^\mathrm{sim}_\mathrm{2D}} = \frac{r^\mathrm{obs}_\mathrm{2D}}{r^\mathrm{sim}_\mathrm{s}}\left(\frac{r_\mathrm{2D}}{r_\mathrm{s}}\right)^{-1}= \frac{20\,\mathrm{pc}}{3.6\,\mathrm{kpc}}\left(\frac{r_\mathrm{2D}}{r_\mathrm{s}}\right)^{-1} = \frac{1}{180}\left(\frac{r_\mathrm{2D}}{r_\mathrm{s}}\right)^{-1}
\end{equation}
using the inner radius, where \(r_\mathrm{2D}/r_\mathrm{s}\) is read off the horizontal axis of \cref{fig:VegettiRatio}. To match both mass observations as closely as possible, we minimized
\begin{equation}\label{eq:chi2}
\begin{split}
    \chi^2 = &\left[\left(\frac{\mu M(r < r^\mathrm{sim}_\mathrm{2D}) - M(r < 20\,\mathrm{pc})}{\sigma(M(r < 20\,\mathrm{pc}))}\right)^2\right.\\ 
    &+ \left.\left(\frac{\mu M(r < 4.5r^\mathrm{sim}_\mathrm{2D}) - M(r < 90\,\mathrm{pc})}{\sigma(M(r<90\,\mathrm{pc}))}\right)^2\right]
\end{split}
\end{equation}
with the respective uncertainties, \(\sigma\). This yields
\begin{equation}
\begin{split}
    \mu = &\left(\frac{M(r < r^\mathrm{sim}_\mathrm{2D})M(r < 20\,\mathrm{pc})}{\sigma^2(M(r < 20\,\mathrm{pc}))} + \frac{M(r < 4.5r^\mathrm{sim}_\mathrm{2D})M(r < 90\,\mathrm{pc})}{\sigma^2(M(r < 90\,\mathrm{pc}))}\right)\\
    &\times\left(\frac{M^2(r < r^\mathrm{sim}_\mathrm{2D})}{\sigma^2(M(r < 20\,\mathrm{pc}))} + \frac{M^2(r < 4.5r^\mathrm{sim}_\mathrm{2D})}{\sigma^2(M(r < 90\,\mathrm{pc}))}\right)^{-1}.
\end{split}
\end{equation}
The resulting \(\mu\) and \(\lambda\) fully specify the rescaled halo, its cross section, and evolution time, as listed in \cref{table:RescaledQuantities}.

\end{appendix}
\end{document}